\documentclass[]{aastex631}

\usepackage[libertine]{newtxmath}
\usepackage[outercaption]{sidecap}    
\usepackage{longtable}

\newcommand{\rulesep}{\unskip\ \vrule\ }

\shorttitle{Debiasing MITHNEOS}
\shortauthors{Marsset et al.}
\graphicspath{{./}{figures/}}
\RequirePackage[normalem]{ulem} 
\RequirePackage{color}\definecolor{RED}{rgb}{1,0,0}\definecolor{BLUE}{rgb}{0,0,1} 
\providecommand{\DIFaddbegin}{} 
\providecommand{\DIFaddend}{} 
\providecommand{\DIFdelend}{} 

\begin{document}

\title{The debiased compositional distribution of MITHNEOS:\\ Global match between the near-Earth and main-belt asteroid populations and excess of D-type NEOs}

\correspondingauthor{Micha\"el Marsset}
\email{mmarsset@mit.edu}

\author[0000-0001-8617-2425]{Micha\"el Marsset}
\affil{Department of Earth, Atmospheric and Planetary Sciences, MIT, 77 Massachusetts Avenue, Cambridge, MA 02139, USA}
\affil{European Southern Observatory, Alonso de C\'ordova 3107, Santiago, Chile}

\author[0000-0002-8397-4219]{Francesca E. DeMeo}
\affiliation{Department of Earth, Atmospheric and Planetary Sciences, MIT, 77 Massachusetts Avenue, Cambridge, MA 02139, USA}

\author[0000-0002-6423-0716]{Brian Burt}
\affil{Lowell Observatory, 1400 W. Mars Hill Road, Flagstaff, AZ, 86001, USA}

\author[0000-0002-6977-3146]{David Polishook}
\affiliation{Faculty of Physics, Weizmann Institute of Science, Israel}

\author[0000-0002-9995-7341]{Richard P. Binzel}
\affiliation{Department of Earth, Atmospheric and Planetary Sciences, MIT, 77 Massachusetts Avenue, Cambridge, MA 02139, USA}

\author[0000-0002-5624-1888]{Mikael Granvik}
\affiliation{Department of Physics, University of Helsinki, PO Box 64, FI-00014 Helsinki, Finland}
\affiliation{Asteroid Engineering Laboratory, Space Systems, Lule\aa{} University of Technology, Box 848, S-981 28 Kiruna, Sweden}

\author[0000-0002-2564-6743]{Pierre Vernazza}
\affiliation{Aix Marseille Univ, CNRS, LAM, Laboratoire d'Astrophysique de Marseille, Marseille, France}

\author[0000-0001-5242-3089]{Benoit Carry}
\affiliation{Universit\'e C{\^o}te d'Azur, Observatoire de la C{\^o}te d'Azur, CNRS, Laboratoire Lagrange, France}

\author[0000-0003-4191-6536]{Schelte J. Bus}
\affiliation{Institute for Astronomy, University of Hawaii, 2860 Woodlawn Drive, Honolulu, HI 96822-1839, USA}

\author[0000-0003-3291-8708]{Stephen M. Slivan}
\affiliation{Department of Earth, Atmospheric and Planetary Sciences, MIT, 77 Massachusetts Avenue, Cambridge, MA 02139, USA}

\author[0000-0003-3091-5757]{Cristina A. Thomas}
\affiliation{Department of Astronomy and Planetary Science, Northern Arizona University, PO Box 6010, Flagstaff, AZ 86005.}

\author[0000-0001-6765-6336]{Nicholas A. Moskovitz}
\affil{Lowell Observatory, 1400 W. Mars Hill Road, Flagstaff, AZ, 86001, USA}

\author[0000-0002-9939-9976]{Andrew S. Rivkin}
\affiliation{Johns Hopkins University Applied Physics Laboratory, Laurel, MD USA}



\begin{abstract}

We report 491 new near-infrared spectroscopic measurements of 420 near-Earth objects (NEOs) collected on the NASA InfraRed Telescope Facility (IRTF) as part of the MIT-Hawaii NEO Spectroscopic Survey (MITHNEOS). 
These measurements were combined with previously published data \citep{Binzel:2019} and 
bias-corrected to derive the intrinsic compositional distribution of the overall NEO population, as well as of subpopulations coming from various escape routes (ERs) in the asteroid belt and beyond. 
The resulting distributions reflect well the overall compositional gradient of the asteroid belt, with decreasing fractions of silicate-rich (S- and Q-type) bodies and increasing fractions of carbonaceous (B-, C-, D- and P-type) bodies as a function of increasing ER distance from the Sun. 
The close compositional match between NEOs and their predicted source populations validates dynamical models used to identify ERs and argues against any strong composition change with size in the asteroid belt between $\sim$5\,km down to $\sim$100\,m. 
A notable exception comes from the over-abundance of D-type NEOs from the 5:2J and, to a lesser extend, the 3:1J and $\upnu_6$ ERs, hinting at the presence of a large population of small D-type asteroids in the main belt. 
Alternatively, this excess may indicate preferential spectral evolution from D-type surfaces to C- and P-types as a consequence of space weathering, or to the fact that D-type objects fragment more often than other spectral types in the NEO space. 
No further evidence for the existence of collisional families in the main belt, below the detection limit of current main-belt surveys, was found in this work.

\end{abstract}

\section{Introduction} \label{sec:intro}

Owing to its close proximity to the Earth, the population of Near-Earth Objects (NEOs) represents both a threat to civilization by its potential for catastrophic impacts, as well as attractive targets for space exploration and resource utilization. 
Forming the bridge between meteorites found on Earth and the parent populations of planetesimals from which these objects originated, 
NEOs are also highly valuable from a purely scientific point of view. 
Matching them to meteorites and escape regions (or escape routes, ERs) in the asteroid belt and beyond will ultimately provide a better understanding of the original chemical and thermal gradient of the young Solar System.  

Linking NEOs to their ERs first requires the use of dynamical numerical codes 
simulating the evolution of asteroid orbits from the main belt onto near-Earth orbits. 
The mechanism by which a small asteroid enters the NEO space starts with its migration 
in the asteroid belt due to the Yarkovsky effect \citep{Bottke:2006,Vokrouhlicky:2015,Granvik:2017}, 
according to which uneven diurnal heating and cooling of a small (D$<$40\,km) body's surface produces a drift of its orbit. 
When the object reaches a major secular or mean-motion resonance (MMR) with Jupiter or Saturn, 
its orbital eccentricity quickly increases until the object is scattered out from the belt \citep{Wetherill:1979, Wisdom:1983}, possibly onto an Earth-crossing orbit, 
where it will remain for typical timescales of a few million years \citep{Bottke:2002,Morbidelli:2002,Granvik:2018}. 

From a dynamical perspective, the secular $\upnu_6$ resonance delimiting the inner edge of the main belt, and the 3:1 MMR 
with Jupiter, separating the inner from the middle belt, were found to be the most effective pathways to NEO space \citep{Bottke:2000,Morbidelli:2002}. 
This was confirmed by observations of 
spectral similarities between NEOs and main-belt asteroids populating the inner belt (e.g., \citealt{Vernazza:2008, Carry:2016}). 
However, other ERs, including outer main-belt resonances and 
the Jupiter-family comets (JFCs), 
also provide non-negligible contributions to the total NEO population \citep{Fernandez:2001,Fernandez:2005,Bottke:2002,Morbidelli:2002,Demeo:2008}. 

Surface characterization is another key component for advancing knowledge about the origins of NEOs and, therefore, meteorites. 
Large-scale surveys conducted in visible and infrared wavelengths provide today a great view on the compositional distribution of NEOs \citep{Ivezic:2001, Binzel:2004, Binzel:2019, Mainzer:2011, Demeo:2013, Demeo:2014, Carry:2016, Perna:2018, Devogele:2019, Sergeyev:2021} and 
inform us about the physical processes 
acting on NEO surfaces \citep{Chapman:1996, Chapman:2004, Sasaki:2001, Binzel:2004, Binzel:2010, Binzel:2015, Moroz:2004, Brunetto:2005, Brunetto:2006, Brunetto:2014, Nesvorny:2005, Nesvorny:2010, Strazzulla:2005, Marchi:2006, Lazzarin:2006, Noble:2007, Vernazza:2009, Nakamura:2011, Noguchi:2011, Delbo:2014, DeMeo:2014_Mars, Polishook:2014, Lantz:2017, Graves:2018, Graves:2019}. 
In that context, the MIT-Hawaii Near-Earth Object Spectroscopic Survey (MITHNEOS) has been acquiring near-infrared (NIR) spectroscopic observations of NEOs on the NASA Infrared Telescope Facility (IRTF) for nearly two decades. 
Recent release of several hundred spectra acquired up until December 2014 highlighted, among other results, the spectral diversity of NEOs, their lack of compositional change over two orders of magnitude in size (10\,km to 100\,m), as well as their preferred ERs as a function of compositional classes \citep{Binzel:2019}.
Several problems however currently persist in our comprehension of the NEO population, including the current 
underrepresentation of carbonaceous (C-type) NEOs compared to the inner belt \citep{Binzel:2015, Binzel:2019}. 

In this paper, we 
focus on the spectral characterization of NEOs using measurements from MITHNEOS. 
This work extends the work performed by \citet{Binzel:2019} by bias-correcting the sample and including 491 additional spectra of 420 individual NEOs that have been collected in the frame of this program after 2014 but not yet published. 
Armed with this new dataset, we compare the compositional distribution of NEOs to their predicted source populations in the asteroid belt to: 
(1) assess the validity of numerical codes used to investigate the origins and evolution of NEOs, 
(2) investigate compositional change with size in the asteroid belt, below the current characterization limit of main-belt surveys and 
(3) search for compositional differences that may provide new insight to the mechanisms responsible for the delivery of extraterrestrial material to the Earth. 

Our paper is organized as follows: 
in Section~\ref{sec:observations}, we present our observing strategy and the taxonomic classification of our new spectra. 
In Section~\ref{sec:debias}, we present the method used to debias our dataset. 
In Section~\ref{sec:results}, we present the debiased NEO population and compare it to the main asteroid belt. 
Finally, we summarize our work in Section~\ref{sec:summary}. 




\section{Data acquisition and classification}
\label{sec:observations}

\subsection{Observations}

\begin{figure}[h!]
\centering
\includegraphics[angle=0, width=0.49\linewidth, trim=0cm 0cm 0cm 0cm, clip]{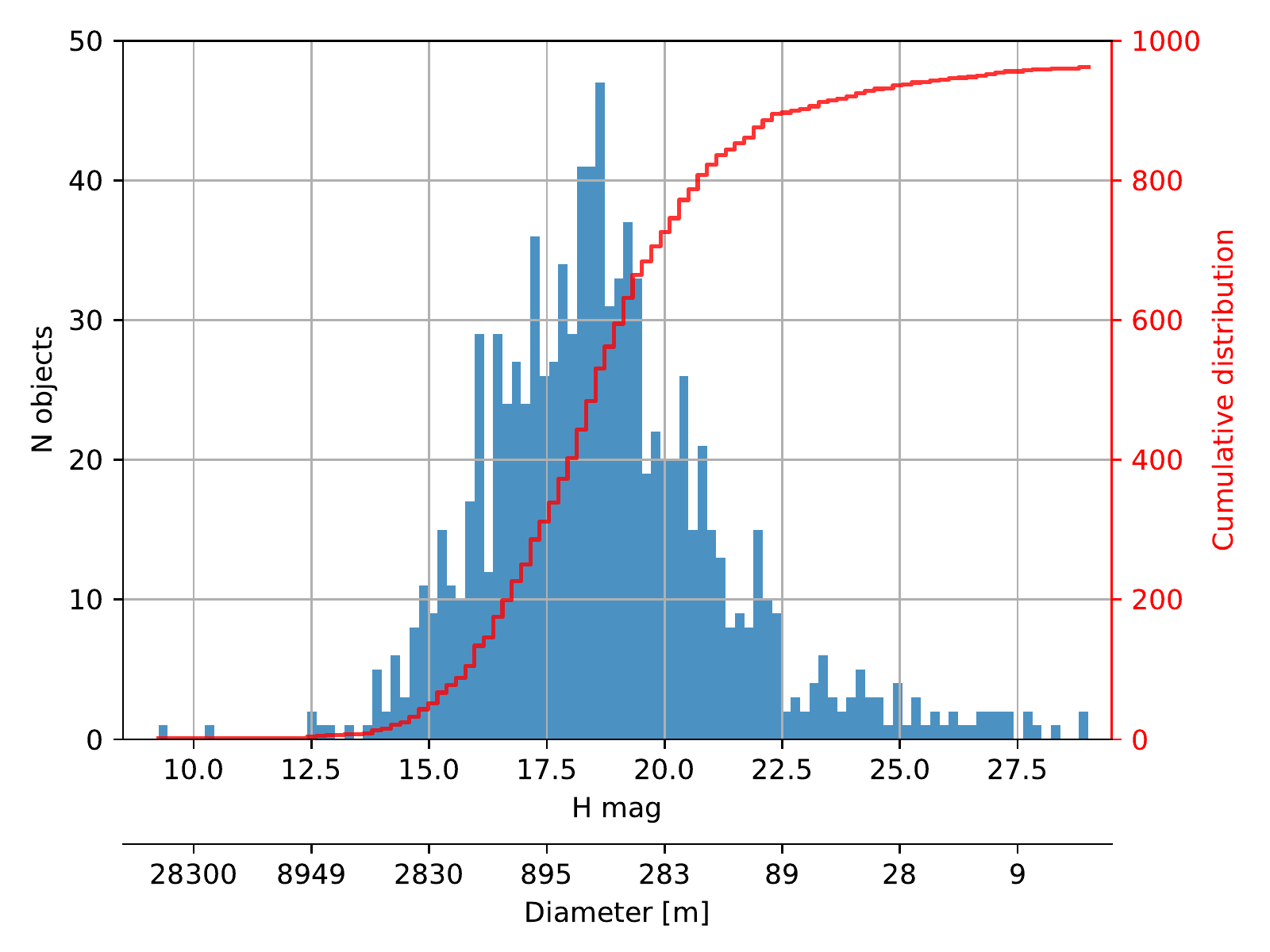}
 \caption{Histogram of absolute ($H$) magnitude and diameter of the MITHNEOS dataset of 976 NEOs ($n_{bin}=100$). $H$ magnitude values were converted to diameters using the measured average albedo $p_v=0.22$ of the dataset. Note the small-size (large-$H$) end tail of the distribution: this reflects the effort made to observe as many small NEOs as possible in the frame of the survey.}
\label{fig:histo_size}
\end{figure}

We present 491 new NIR spectra of 420 NEOs mainly between 100\,m to 3\,km in diameter (Fig.\,\ref{fig:histo_size}). 
These observations were collected for the most part between January 2015 and February 2021 through the MITHNEOS program, 
with the exception of 10 spectra acquired earlier (between 2003 and 2013), but inadvertently never published before. 
Combined with our previously published spectra \citep{Binzel:2019}, this leads to a dataset of NIR spectral measurements for a total of 976 NEOs (many with multiple measurements available) that we analyse in this work.  

Spectroscopic observations were conducted with the 3-meter IRTF located on Mauna Kea, Hawaii. 
We used the SpeX NIR spectrograph \citep{Rayner:2003} combined with a 0.8$\times$15 arcsec slit in the low-resolution prism mode to measure the spectra over the 0.7--2.5~micron wavelength range. 
Series of spectral images with 120\,s exposure time were recorded in an AB beam pattern to allow efficient removal of the sky background by subtracting pairs of AB images. 
Asteroid observations were alternated with measurements of calibration stars known to be very close spectral analogs to the Sun: Hyades 64 and \citet{Landolt:1983}'s stars 93-101, 98-978, 102-1081,105-56, 107-684, 107-998, 110-361, 112-1333, 113-276 and 115-271. 
We typically observe three different stars each night 
to be able to identify possible outlier measurements and to mitigate spectral variability across the observations by computing a mean stellar spectrum from the three measurements. 
An in-depth analysis of these calibration stars is provided in \citet{Marsset:2020_Spex}. 

Data reduction and spectral extraction followed the procedure outlined in \citet{Binzel:2019}. 
We summarize it briefly here. 
Reduction of the spectral images was performed with the Image Reduction and Analysis Facility (IRAF) and Interactive Data Language (IDL), using the Autospex software tool to automatically write sets of command files \citep{Rivkin:2005Icar..175..175R}. 
Reduction steps for the science targets and their corresponding calibration stars included trimming the images, creating a bad pixel map, flat fielding the images, sky subtracting between AB image pairs, tracing the spectra in both the wavelength and spatial dimensions, co-adding the spectral images, extracting the spectra, performing wavelength calibration, and correcting for air mass differences between the asteroids and the corresponding solar analogs. 
Finally, the resulting asteroid spectra were divided by the mean stellar spectra to remove the solar gradient. 

All reduced spectra collected through MITHNEOS and information about their observing conditions are made rapidly available to the community by being posted online at \url{http://smass.mit.edu/minus.html}. 
Thumbnail figures of the new spectral dataset are presented in Appendix\,\ref{sec:thumbnails} (the complete figure set is available only in the online journal). 
The corresponding list of spectra, including results of our spectral analysis, information about the NEOs's physical and orbital properties, observing conditions, and ER probabilities (Section\,\ref{sec:HFD}) is provided in Table~\ref{tab:suppmat} in Appendix~\ref{sec:app_E}.

\subsection{Taxonomic assignment}
\label{sec:taxons}

Taxonomic classification of our new dataset was performed as follows. For any spectra for which a thermal tail is detected, a correction to remove the thermal component was applied before classification as described in following Section~\ref{sec:thermal}. 
Spectra were then run through the Bus-DeMeo Principal Component Analysis (PCA) online classification tool \citep{DeMeo:2009} developed by Stephen M. Slivan and publicly available at \url{http://smass.mit.edu/busdemeoclass.html}. We refer the readers to \citet{Tholen:1984}, \citet{Tholen:1989} and \citet{Bus:1999tw} for explanations of the PCA method applied to asteroid spectroscopy. 

Spectra that have both visible and near-infrared data available are typically assigned a unique class through this method. Spectra with only near-infrared data, which includes the majority of our data, are typically assigned multiple possible classes through the tool and the data are either degenerate (cannot be assigned a unique class because insufficient wavelength coverage exists) or require visual inspection to identify features that indicate the class. 
Spectra in the S-complex (S-, Sq-,  Sr-, Sv- and Q-types) that are not assigned a unique class with the online tool were then classified according to their 1-$\upmu$m band spectral properties using \citet{DeMeo:2014_Mars}’s decision tree (see their Fig. 3). 

The classes of the C and X complexes (B-, C-, Cb-, Cg-, Cgh-, Ch-, X-, Xc-, Xe-, Xk- and Xn-type) are generally distinguished with features at visible wavelength ranges, so spectra with only near-infrared data are assigned “C,X” in most cases. Three exceptions are Xn- and Xk-types that display a 0.9-micron feature 
and C-types that have a clear broad feature centered near $\sim$1.3. 
Additional slope criteria were also used to help distinguish classes. “C,X” spectra with negative spectral slopes were assigned a B-type, those with a near-infrared slope greater than or equal to 0.20 or 0.38\,${\rm \upmu m^{-1}}$ were assigned X-type and D-type, respectively.

A final visual inspection was performed for all spectra to ensure classifications were sensible. As part of the visual inspection process, we append the notation “:” to the spectral class to indicate uncertainty  when a spectrum does not clearly, unambiguously fall within that class. Objects with uncertain taxonomic assignment due to a low signal-to-noise ratio (SNR) are marked with the notation “::”. Some spectra were assigned the value “U” and are considered Unusual and Unclassified as they do not acceptably fall within any defined class. These objects include 143992, 302830, 453707, and 2012\,TM$_{139}$. Spectra marked with “::” or classified as “U” were not used in our analysis.
Final classifications are provided in Table~\ref{tab:suppmat} in Appendix~\ref{sec:app_E}.

This new classified dataset was then combined with previously classified spectra from MITHNEOS \citep{Binzel:2019}. In the case where an asteroid was observed multiple times and had distinct taxonomic types assigned to its spectra, the most frequently assigned taxonomic type was selected for that asteroid (for instance, if an asteroid was observed three times and two of its spectra were classified as Sq whereas the third one was classified as Sw, then the final classification was Sq). If an asteroid had an equal number of two different taxonomic types assigned to its spectra, then the spectrum with the highest signal-to-noise ratio was used for final classification. 

\subsection{Thermal correction}
\label{sec:thermal}

At Earth's distance from the Sun, an atmosphere-less body with a strongly absorbing surface (low albedo) can reach equilibrium temperature of $\sim$300\,K, i.e., 
high enough to emit detectable amounts of thermal flux at near-infrared wavelengths (e.g., \citealt{Lebofsky:1989, Abell:2003, Rivkin:2005Icar..175..175R, Reddy:2009, Reddy:2012, Binzel:2019}). 
We report here the detection of 17 new objects with an excess near-infrared thermal flux in our dataset. 
In order to remove thermal excess prior to taxonomic assignment, we applied the near-Earth asteroid thermal model (NEATM) developed by \citet{Harris:1998} using a model by \citet{Volquardsen:2007} and implemented by \citet{Moskovitz:2017} as described in \citet{Binzel:2019}. We adopt the same fixed parameter values of those works: emissivity of 0.9, slope parameter G=0.15, and thermal infrared beaming parameter $\eta$ according to the equation in \citet{Masiero:2011} based on phase angle at the time of observation.
Appendix\,\ref{sec:app_C} provides the $\eta$ value used in the model and resulting albedo.

\section{Compositional distribution debiasing}
\label{sec:debias}

\subsection{Method}

Our aim is to derive the bias-corrected compositional distribution of the overall NEO population, as well as of populations coming from individual ERs in the asteroid belt and beyond. 
One of the main bias affecting surveys of small bodies in the Solar System is the one that favors the discovery and characterization of bright objects. 
When observed at visible and NIR wavelengths, for similar sizes, high-albedo asteroids reflect more light from the Sun and therefore can be more easily discovered and characterized than the low-albedo ones.
This leads to an observational preference for classes of high-albedo bodies such as the A-, Q-, S-, E-, M-, and V-types compared to classes of dark objects such as the C-, P- and D-types. 

In order to quantify this effect, let's consider a population 
of NEOs with albedos comprised between $p_v$ and $p_v+dp_v$, sizes between $D$ and $D+dD$, and heliocentric distances between $r$ and $r+dr$. 
At a given location on the sky, the number of objects is given by

\begin{equation} 
n(r,D,p_v) = K\, f(D)\, g(r)\, h(p_v)\, dp_v\, dr\, dD,
\label{eq:eq1}
\end{equation}

where $f(D)$ is the size distribution of the NEO population, g(r) its radial distribution, h($p_v$) its albedo distribution, and $K$ is a normalization constant to obtain the desired number density unit (e.g., number of objects per square degree). 

Ignoring phase effects, the apparent magnitude of an object relates to its diameter $D$ and albedo $p_v$ in the following way:

\begin{equation} 
m = C - 5\,{\rm log_{10}}(D) - 2.5\,{\rm log_{10}}(p_v) + 5\,{\rm log}(r \Delta),
\label{eq:eq2}
\end{equation}

where $\Delta$ is the geocentric range of the object and C is a constant that defines the absolute magnitude system. 

Writing the size D in terms of absolute magnitude m and its derivative with respect to $m$ using equation~\ref{eq:eq2}, 
and assuming that the size distribution of the population can be approximated by a power law of the form $f(D) = A D^{-\alpha}$, where A is a normalization constant and $\alpha$ is the power-law slope, 
equation \ref{eq:eq1} can be rewritten 
in terms of $p_v$ and $m$ as

\begin{equation} 
n(r,D,p_v) = \frac{-{\rm ln}(10) K A}{5} \, h(p_v)\,  p_v^{\frac{\alpha-1}{2}}\, g(r)\, r^{1-\alpha} \Delta^{1-\alpha} 10^{\frac{(1-\alpha)\,(C-m)}{5}} dp_v\, dm\, dr.
\label{eq:eq6}
\end{equation}

For an homogeneous population of NEOs named ``$x$'' all sharing the same albedo (meaning that $h_x(p_v) = \delta(p_{v,x} - p_v)$, where $\delta$ is the Dirac delta function), and comprised between distances $r_0 \leq r \leq r_1$, the number of objects with magnitudes between $m_0$ and $m_1$ is

\begin{equation} 
N_x = \frac{-{\rm ln}(10) K_x A}{5}\, p_{v,x}^{\frac{\alpha-1}{2}}\, \int_{r_0}^{r_1} \int_{m_0}^{m_1} g(r)\, r^{1-\alpha} \Delta^{1-\alpha} 10^{\frac{(1-\alpha)\,(C-m)}{5}} dm\, dr. \\
\label{eq:eq7}
\end{equation}

Therefore, the number ratio $R_{obs}$ of {\it observed} objects from two distinct subpopulations (``$x$'' and ``$y$'') with similar size distribution and radial distributions, but distinct albedos and overall number densities would be






\begin{equation}
R_{obs} (<m) = \frac{N_{obs, x}}{N_{obs, y}}(<m) = \frac{K_{x}}{K_{y}}(<m) \times \,\left (\frac{p_{v, x}}{p_{v, y}} \right)^{2.5 \beta}.
\label{eq:eq11}
\end{equation}

where we substituted $\beta = \frac{\alpha -1}{5}$, and where $\frac{K_{x}}{K_{y}} (<m) = R_{true} (<m)$ represents the true {\it intrinsic} magnitude-limited ratio of objects from the two subpopulations of NEOs. 


The albedo bias affecting the detected population of objects can therefore be accounted for by using an estimate of the average albedo of the taxonomic classes, and knowledge about the power-law slope $\beta$ of the absolute $H$ magnitude frequency distribution (HFD) of the population. This method was previously used, e.g., by \cite{Stuart:2004} for the NEO population, and by \citet{Marsset:2019} and \citet{Schwamb:2019}  in the case of Trans-Neptunian Objects. 

Two main improvements are made here compared to the work of \cite{Stuart:2004}. 
First, we take advantage of the increased number of spectroscopic (this work) and albedo (e.g., \citealt{Mainzer:2011}) measurements to better constrain the average albedo of each taxonomic class of NEOs. 
Second, 
we make use of the numerical model of \cite{Granvik:2018} to probabilistically link each object in our dataset to specific ERs in the asteroid belt, thereby allowing a compositional investigation of individual ERs. 


We stress that equation\,\ref{eq:eq11} was derived under a number of simplifying assumptions and ignoring physical effects, such as differential phase darkening (e.g., \citealt{Luu:1989, Sanchez:2012, Perna:2018, Mahlke:2021}), that could further bias our sample. 
Additional biases inherent to targeted small body surveys are discussed by \citet{Devogele:2019} (see their Section 5). 
Despite ignoring these effects, we show in Section~\ref{sec:NEOs_vs_MBAs} that our debiasing technique allows to account for most of the composition difference between NEOs and their predicted ERs in the asteroid belt, confirming that albedo variation is the dominant surface property affecting NEO discovery and characterization surveys.

\subsection{Albedos}
\label{sec:albedos}

The first set of observables needed to debias the compositional distribution of NEOs by use of equation~\ref{eq:eq11} is the set of albedos of the different classes of objects in our dataset. 
Albedo measurements for these objects were retrieved from the following catalogs and papers: \citet{Delbo:2003}, IRAS \citep{Ryan:2010}, ExploreNEOs \citep{Trilling:2010}, AKARI \citep{Usui:2011}, NEOWISE \citep{Mainzer:2011, Nugent:2016, Masiero:2017, Masiero:2018}, and NEOSurvey \citep{Trilling:2016}. 
For each object, an average value was calculated whenever multiple measurements were available, weighting by the inverse of the squared uncertainty on the measurements. 

Our classified spectra (Section~\ref{sec:taxons}) were then grouped into the following broad taxonomic classes: A-, B-, C-, D-, Q-, S-, K-, L-, V-, and X-type, and a median albedo was calculated for each class. 
The X-complex was further subdivided into E-, M- and P-type \citep{Tholen:1984} based on albedo values, whenever available. Specifically, an object was classified as P-type if $p_v \leq 7.5\%$, M-type if $7.5\% < p_v \leq 30\%$ and E-type if $p_v > 30\%$.
X-complex objects without any albedo measurement available were attributed 
a P-type or an M-type randomly, weighting the probabilities of the spectral type assignment by the P-to-M ratio measured in the asteroid belt \citep{Demeo:2013}. 
Median albedos of the taxonomic classes are reported in Table\,\ref{tab:albedos}. 
Asymmetric uncertainties were calculated by 1-sigma clipping the distribution of albedo values in each class, and then measuring the one-sided root mean square of the remaining distribution in both directions. 
Due to the low number of albedo measurements, no reliable uncertainties could be measured for the A and K classes. 

Because the number of NEOs with measured albedos was low for some taxonomic classes in our dataset, we further used the WISE catalog of Main-belt Asteroids (MBAs) \citep{Masiero:2011} to derive albedo values for the subsample of taxonomically-classified MBAs from the photometric Sloan Digital Sky Survey (SDSS) \citep{Demeo:2013}. 
This dataset provides one order to two orders of magnitude more classified objects than our NEO dataset, with the exception of the class of spectrally ``fresh'' Q-type asteroids more commonly found among NEOs 
(e.g., \citealt{Binzel:2010}). 
Again, the derived median albedo of each class and its associated uncertainties are reported in Table\,\ref{tab:albedos}. 
The values are globally consistent with the ones measured for the NEO sample. 
We note, however, that aside from the B-types, carbonaceous bodies (C-, P- and D-types) have lower albedos in the NEO population than in the asteroid belt. 
This could be due to a change in the optical properties of these bodies as they get closer to the Sun, to NEO observations being performed at higher phase angles on average \citep{Perna:2018}, or simply to a sample size effect. 
While we acknowledge that asteroid surfaces can be spectrally altered as they evolve in the NEO space, our analysis uses the taxonomic albedo values derived from the main-belt dataset, which we consider more reliable because they are based on a much larger number of measurements. 

\begin{deluxetable}{lcccc}[h!]
\tabletypesize{ \scriptsize}
\tablecaption{ Median albedo of taxonomic classes for NEOs and MBAs.}
\tablehead{ & \multicolumn{2}{c}{NEOs (MITHNEOS)} & \multicolumn{2}{c}{MBAs \citep{Demeo:2013}} \\
Taxon & $p_v$ & $N_{obj}^*$ & $p_v$ & $N_{obj}^*$}
\startdata
A & --                     &          2 & 0.26$\pm$0.05          &         33 \\
B & $0.09^{+0.05}_{-0.04}$ &         16 & 0.06$^{+0.02}_{-0.01}$ &        866 \\
C & $0.04^{+0.04}_{-0.01}$ &         13 & 0.06$^{+0.03}_{-0.02}$ &       5018 \\
D & $0.04^{+0.03}_{-0.02}$ &          9 & 0.08$^{+0.03}_{-0.02}$ &        428 \\
Q & $0.25^{+0.07}_{-0.04}$ &         70 & 0.23$^{+0.06}_{-0.05}$ &         46 \\
S & $0.24^{+0.04}_{-0.05}$ &        256 & 0.25$\pm0.05$          &       6697 \\
K & --                     &          4 & 0.16$\pm0.05$          &         902 \\
L & 0.12$\pm$0.04          &         17 & 0.17$^{+0.04}_{-0.05}$ &         692 \\
V & $0.34^{+0.08}_{-0.07}$ &         16 & 0.35$\pm$0.06          &        718 \\
E & $0.55^{+0.07}_{-0.12}$ &         23 & 0.51$^{+0.11}_{-0.18}$ &         43 \\
M & 0.15$\pm$0.03          &         29 & 0.14$\pm$0.03          &        770 \\
P & 0.03$\pm$0.01          &         26 & 0.05$\pm$0.01          &        752 \\
\enddata
\tablenotetext{}{$^*$Number of objects used to compute the median albedo value of the taxonomic class.}
\label{tab:albedos}
\end{deluxetable}





\subsection{Escape regions and H magnitude distributions}
\label{sec:HFD}

The second set of parameters needed to debias our sample are the power-law indices $\beta$ of the NEO $H$ magnitude distribution, which are both ER-dependent and magnitude-dependent. 
We retrieved these parameters for the seven ERs explored by \citet{Granvik:2018}, namely (by order of increasing distance to the Sun): 

\begin{itemize}
\item the Hungarias\footnote{To be distinguished from the Hungaria \textit{family}, a collisional family of E-type bodies included in the larger Hungaria \textit{group}.} (Hun) -- a group of high inclination (i$\sim$20${\degr}$) asteroids isolated at the innermost edge of the main belt
(a$<$2.0\,au) by the 4:1 MMR of Jupiter, 
\item the $\upnu_6$ complex, which delimits the inner edge of the main belt, 
\item the Phocaeas (Pho) -- another group of high-inclination (i$\sim$25${\degr}$) bodies isolated inside of the 3:1 MMR (a$<$2.5\,au), 
\item the 3:1, 5:2, and 2:1 complexes of MMRs near 2.5, 2.8, and 3.3~au, respectively, 
\item the JFCs
\end{itemize}

Note that in the Granvik model, the $\upnu_6$, 3:1, 5:2 and 2:1 complexes each regroup several resonances. For instance, the $\upnu_6$ complex includes the $\upnu_6$ secular resonance, as well as the 4:1 and 7:2 MMRs with Jupiter.

$\beta$ parameters were retrieved by fitting the HFDs of each ER by a single power-law function over two absolute magnitude ranges, $H<19$ and $19\,\leq\,H<23$, and measuring the slope of the fitting functions. 
These magnitude ranges were chosen because of (1) the range of magnitudes of the MITHNEOS objects ($\sim$90\% fall within $15<H<23$), and (2) the slope break of the NEO $H$ magnitude distribution near $H=19$ \citep{Granvik:2018}. 




Each NEO in our dataset was then linked to a probability distribution function of ERs (and, therefore, of $H$-magnitude distributions) by use of the model of \citet{Granvik:2018}. 
Specifically, given the orbital elements (a, e, i) and $H$ magnitude of the object as input, we computed its probability distribution function comprised of seven discrete values for each of the seven regions. 
ER probabilities are provided for each new object in our dataset as part of Table~\ref{tab:suppmat} in Appendix~\ref{sec:app_E}. 
The fractional abundances of a given taxonomic group of NEOs in the seven ERs was then obtained by simply dividing its summed ER probabilities by the summed probabilities of the complete set of NEOs. 

One object in our dataset, (3552)~Don Quixote, has a semi-major axis of a=4.26~au falling slightly beyond the range of distances in the model of \citet{Granvik:2018}, which has an outer boundary at 4.2~au. ER probabilities for this object were derived by changing its semi-major axis from 4.26~au to 4.19~au, assuming that the probabilities
do not vary significantly between neighbouring cells of the model.

Our sample also includes 3 active comets. Two of these objects, 162P/Siding Spring (2004 TU12) and 169P/NEAT (2002 EX12) have orbital properties fully consistent with that of Jupiter-Family Comets, while the third one, P/2006 HR30 (Siding Spring), belongs to the family of Halley-type comets. Due its uniqueness in our dataset, P/2006 HR30 was not considered in the rest of our analysis.


For each ER, the compositional distribution was derived separately for $H<19$ and $H>19$, and each magnitude range was debiased independently by use of Equation\,\ref{eq:eq11}, albedo values from Table~\ref{tab:albedos} and $\beta$ slope values from Table~\ref{tab:beta}. 
The debiased fractional abundances from the two magnitude ranges were then summed, weighted by the number of observed objects in each range, and re-normalized to obtain the final bias-corrected compositional distribution of the ER.  
The resulting debiased population is presented and compared to the observed population in Section\,\ref{sec:distributions}.


\begin{deluxetable}{l | c | cc | c | cc}[h!]
\tabletypesize{ \scriptsize}
\tablecaption{ Slope of the $H$ magnitude distributions of NEOs coming from each source region. }
\tablehead{Source region & \multicolumn{2}{c}{Power-law index $\beta$} \\
     & $H<19$ & $19<H<23$}
\startdata
$\upnu_6$    & 0.469  & 0.281  \\
3:1J   & 0.332  & 0.416  \\
Hun    & 0.074  & 0.382  \\
Pho    & 0.239  & 0.054  \\
5:2J   & 0.399  & 0.138  \\
2:1J   & 0.552  & 0.050  \\
JFC    & 0.125  & 0.348  \\
\enddata
\tablenotetext{}{}
\label{tab:beta}
\end{deluxetable}

\subsection{Main-belt comparison dataset}
\label{sec:criteria}

In an attempt to validate NEO migration models, the main question we aim to address is the following: \textit{does the compositional distribution of NEOs match their predicted dynamical source populations in the asteroid belt?} The answer will also help us investigate composition change with size in the asteroid belt. 

In order to obtain representative samples of the source populations feeding each ER in the main belt, we used the spectrally-classified SDSS dataset  
of \cite{Demeo:2013}. 
The first step consisted in reexamining the size completeness limit of this dataset as a function of heliocentric distance and spectral class. 
To do so, it is a common practice 
to fit the size frequency distribution (SFD) of the dataset   
by a single power law, and then by measuring the size at which the SFD deviates from the power law.  
The issue with this method stems from the fact that distinct composition classes with distinct albedos necessarily have distinct size completeness limits. 
The completeness limit of a compositionally mixed population will necessarily lead to an underestimation of the number of dark bodies in the population and an overestimation of the number of bright objects. 

\begin{figure}[h!]
\centering
\includegraphics[angle=0, width=0.325\linewidth, trim=0.5cm 0cm 0.75cm 0cm, clip]{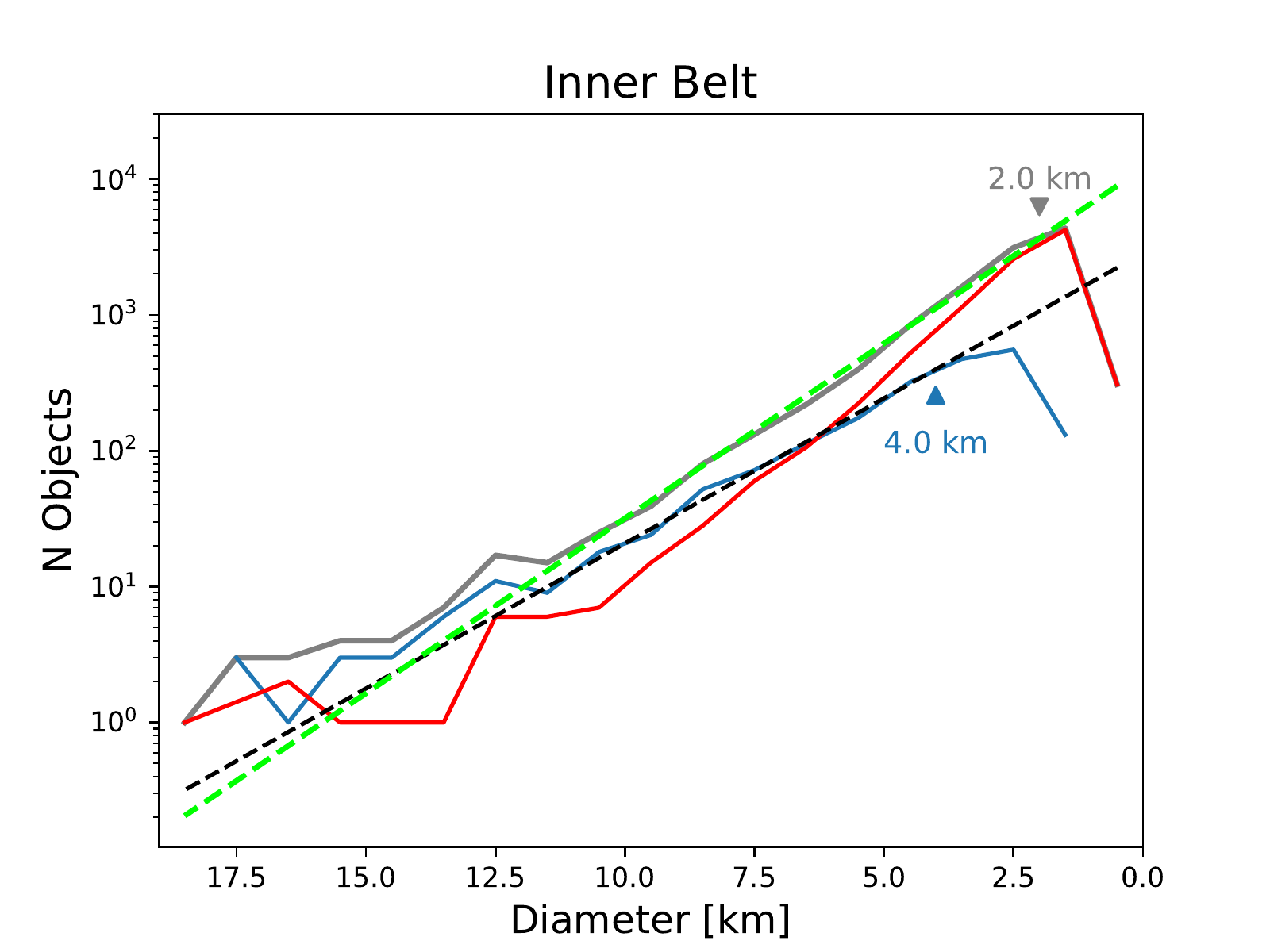}
\includegraphics[angle=0, width=0.325\linewidth, trim=0.5cm 0cm 0.75cm 0cm, clip]{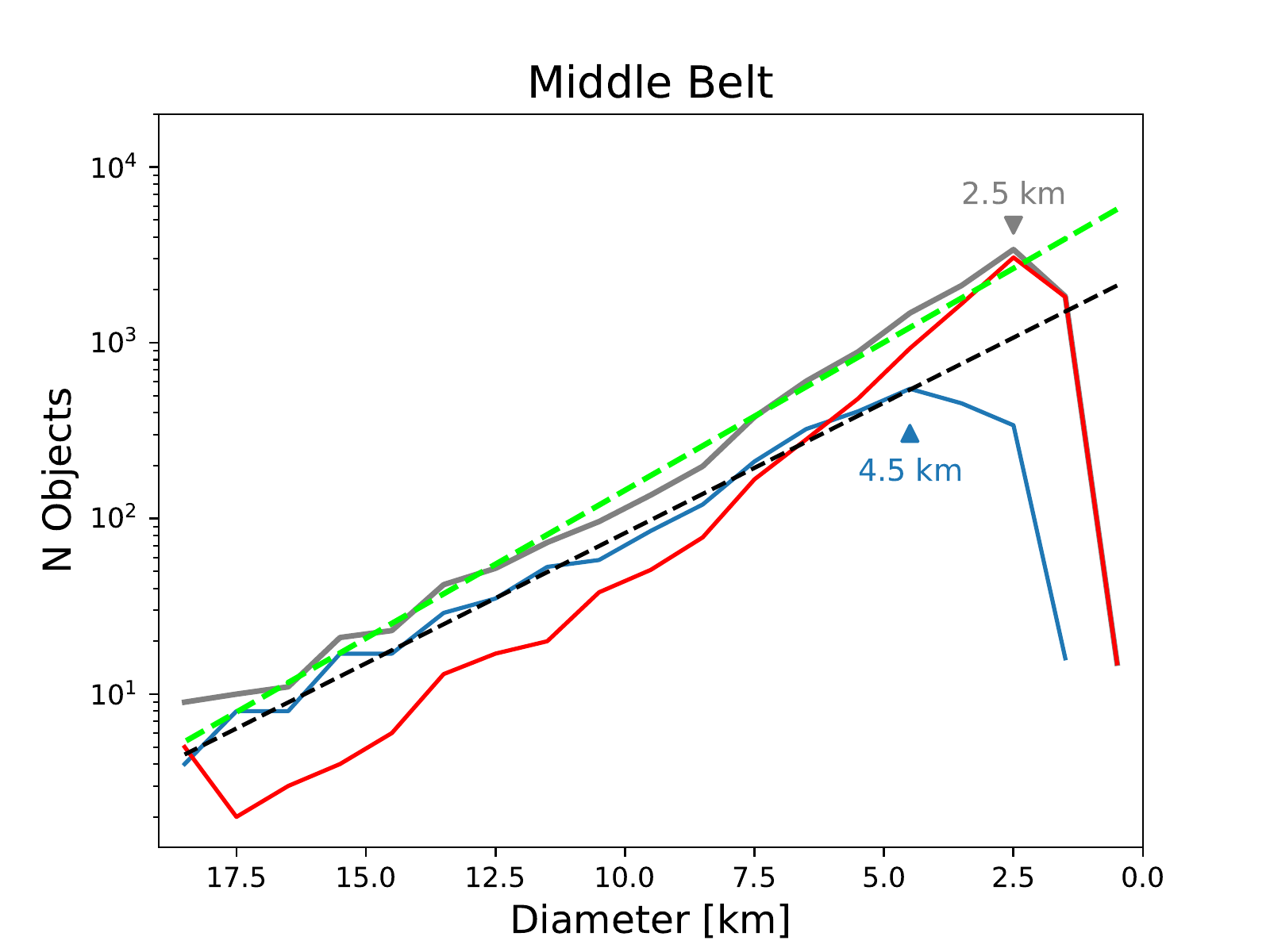}
\includegraphics[angle=0, width=0.325\linewidth, trim=0.5cm 0cm 0.75cm 0cm, clip]{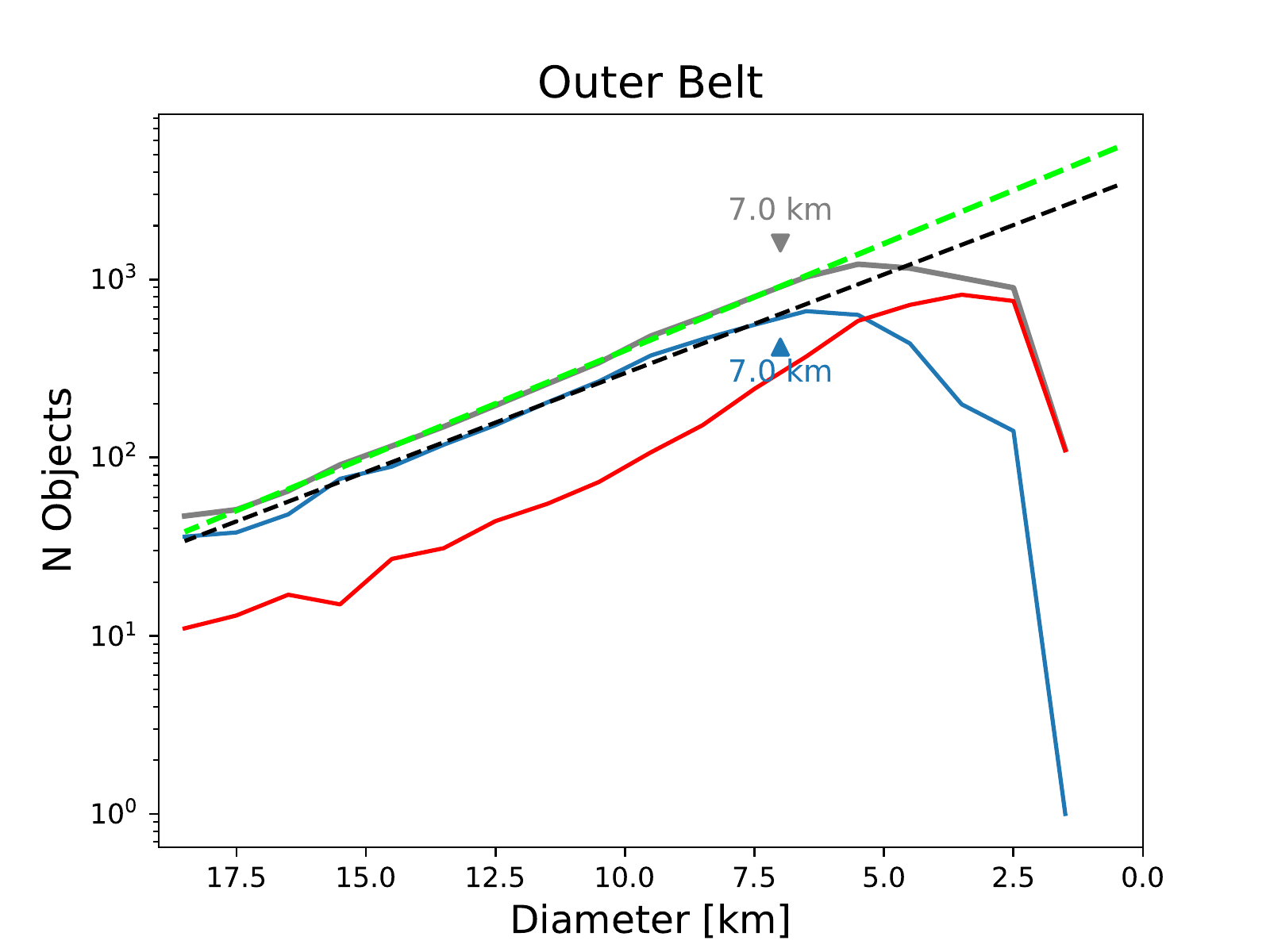}
\includegraphics[angle=0, width=0.325\linewidth, trim=0.5cm 1cm 0.75cm 7.5cm, clip]{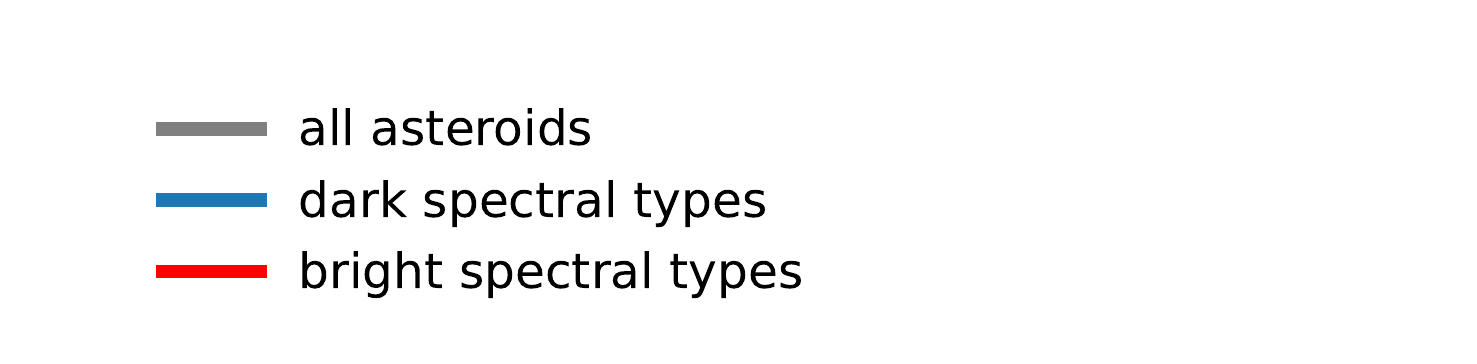}
 \caption{Number of main-belt asteroids selected by \citet{Demeo:2013} from the SDSS moving object catalog, plotted as a function of asteroid diameter (bin size of 0.5 km) and region of the belt ({\it left:} inner belt, {\it center:} middle belt, {\it right:} outer belt). The completeness limit of three regions is determined by fitting a power-law function to the size distribution and measuring the size at which the distribution diverges from the power law. The fitting power-law functions to the full population and to the subset of dark spectral classes are represented by the dashed green and black lines, respectively. The fitting function to the subset of bright classes is not shown for better readibility. Deriving size completeness limits using the full, compositionally-mixed population or using the bright subset leads to an artificially biased distribution against dark objects (Section~\ref{sec:criteria}). Therefore, we only use the dark subset and derive size completeness limits of 4.0\,km for the inner belt, 4.5\,km for the middle belt, and 7.0\,km for the outer belt.} 
\label{fig:SDSS_SFD}
\end{figure}

This effect is illustrated in Fig.~\ref{fig:SDSS_SFD},  where we compared the SFD of the 
taxonomically-classified dataset of \citet{Demeo:2013} to the SFDs of its dark (B-, C-, P-, and D-type asteroids) and bright (A-, S-, Q-, K-, L-, V-, E- and M-types) subcomponents. 
It is clear from this figure that the dark subpopulation deviates from the fitted single power law function at a slightly larger diameter than the overall population, and an even larger diameter compared to the bright subpopulation. For instance, in the inner belt, the dark classes deviate at diameter of 4\,km whereas the overall population deviates near diameter of 2\,km. 
While this difference might seem small, it triggers important implications for the resulting composition of the size-limited sample due to the steep SFD of the asteroid belt. 
This steep distribution implies that most objects in a size-limited sample are close to the lower-size boundary and, therefore, to the potentially incorrect completeness limit. 

To mitigate this bias, we adopted a conservative approach consisting in deriving the size completeness limit from the dark subpopulation alone (\citealt{Demeo:2013} had a similar approach that consisted in deriving absolute magnitude cuts for each taxonomic class individually). 
By doing so, we derive size completeness limits of 4\,km for the inner belt (defined as inward of the 3:1J MR, i.e., semi-major axis a$<$2.5\,au), 4.5\,km for the middle belt (between the 3:1J and 5:2J MMR, i.e. 2.5$<$a$<$2.8\,au), and 7.0\,km for the outer belt (beyond the 5:2J, i.e. a$>$2.8\,au; Fig.~\ref{fig:SDSS_SFD}). 
These values constitute the lower-size boundary of our MBA comparison sample. 
The upper boundary was chosen to be 3\,km larger, in order to obtain a large number of objects in our comparison sample while still being in a range of sizes that is not too different from our NEO dataset. 

Next, we built a representative sample of each source population feeding the ERs. This was achieved by measuring the ranges of initial semi-major axis and inclination values of the test particles ending up in the ERs in the dynamical simulation of \citet{Granvik:2018} (see their Figure\,3, second panel), and selecting MBAs falling within these ranges of values. Selection criteria for each source population are summarized in Table~\ref{tab:ERs}. 
Unfortunately, the resulting low number of Hungaria MBAs (22 objects) does not allow any statistically significant comparative analysis between Hungaria NEOs and these objects. Increasing the upper size boundary of the Hungaria comparison sample only adds a few objects to the sample. 


Next, we evaluated the statistical robustness of the resulting taxonomic ratios in our comparison datasets. 
To do so, the lower and upper size limits of the datasets were varied independently over a 1\,km range, with incremental steps of 0.1\,km, in order to measure the variation of the taxonomic ratios as a function of varying size boundaries. 
For instance, in the case of the inner-belt ERs (the $\upnu_6$, 3:1J, and Phocaea), the lower size limit was varied from 4 up to 5\,km while keeping the upper limit fixed to 7\,km,  
then the upper limit was varied from  7 down to 6\,km while keeping the lower limit fixed to 4\,km. 
This resulted in a total of 22 datasets for each ER from which the taxonomic ratio uncertainty linked to size boundary selection, defined as the 1-$\upsigma$ variation of the taxonomic ratios, was measured. 
Final error bars on the taxonomic ratios of the comparison datasets are the quadratic sum of the Poisson uncertainties (i.e., $\sqrt{N}$, where N is the number of objects in the dataset) and the size boundary uncertainties. 

\begin{deluxetable}{ccccc}[h!]
\tabletypesize{ \scriptsize}
\tablecaption{Selection criteria for the main-belt comparison samples. }
\tablehead{Source region & Diameter & Semi-major axis & Inclination & N objects$^*$ \\
& [km] & [au] & [deg] }
\startdata
Hun        & 4.0--7.0  & 1.7--2.1 & $>$12.5 & 22 \\
$\upnu_6$  & 4.0--7.0  & 2.1--2.5 & $<$17.5 & 1407 \\
Pho        & 4.0--7.0  & 2.1--2.5 & $>$17.5 & 72 \\
3:1J       & 4.0--7.0  & 2.2--2.7 & -- & 3412 \\
5:2J       & 4.5--7.5  & 2.6--3.1 & -- & 3225 \\
2:1J       & 7.0--10.0  & 3.0--3.4 & -- & 1596 \\
\enddata
\tablenotetext{}{$^*$Number of objects in the comparison samples, obtained by applying the size and orbital selection criteria. }
\label{tab:ERs}
\end{deluxetable}

\subsection{Additional class merging}
\label{sec:classes}

The MITHNEOS dataset and the SDSS-derived MBA comparison sample are composed of spectroscopic and photometric data collected over two distinct wavelengths ranges: 
the near-infrared range (0.7--2.5~$\upmu$m) for MITHNEOS and 
the visible range (0.3--1.0~$\upmu$m) for SDSS. 
Several taxonomic classes of asteroids can only be distinguished in one of these two wavelength ranges. This is the case, in particular, for the C- and X-complexes that are often indistinguishable in the near-infrared (see Section~\ref{sec:taxons}). 
Similarly, Q- and S-type asteroids can hardly be distinguished at low spectral resolution in visible wavelengths. 

In order to allow a direct comparison of our NEO dataset with the asteroid belt, we therefore merged objects belonging to the C- and X-complexes (more specifically, objects from the C-complex and the low-albedo objects from the X-complex, i.e., the P types) into a single class, hereafter simply named "C/P". 
Similarly, Q-type and S-complex asteroids were merged into a single class ``S/Q''. 

\section{Results}
\label{sec:results}


\subsection{The observed and debiased distributions}
\label{sec:distributions}

\begin{figure}[h!]
\centering
\includegraphics[angle=0, width=0.49\linewidth, trim=5cm 1.5cm 1cm 2cm, clip]{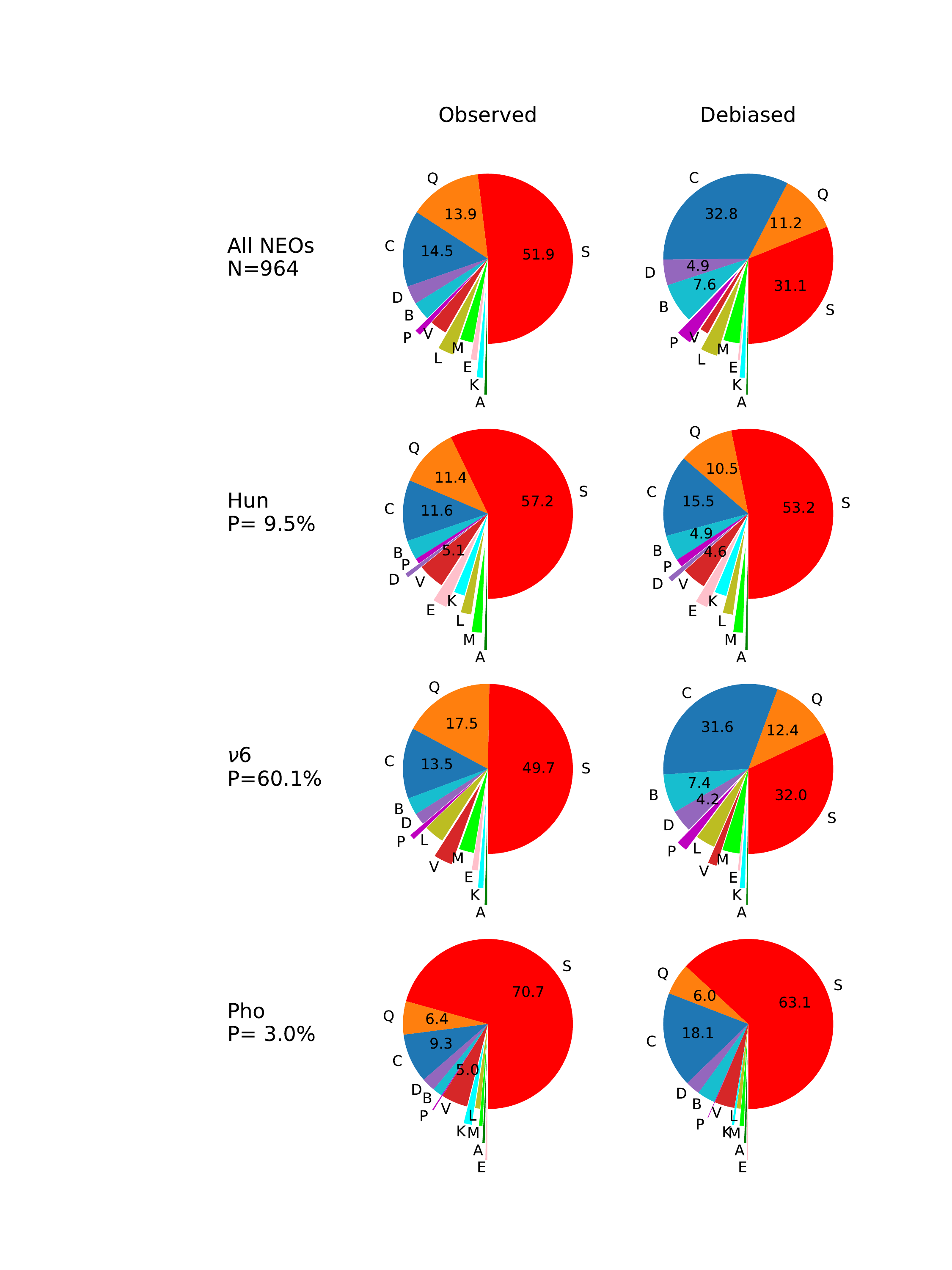}
\rulesep
\includegraphics[angle=0, width=0.49\linewidth, trim=5cm 1.5cm 1cm 2cm, clip]{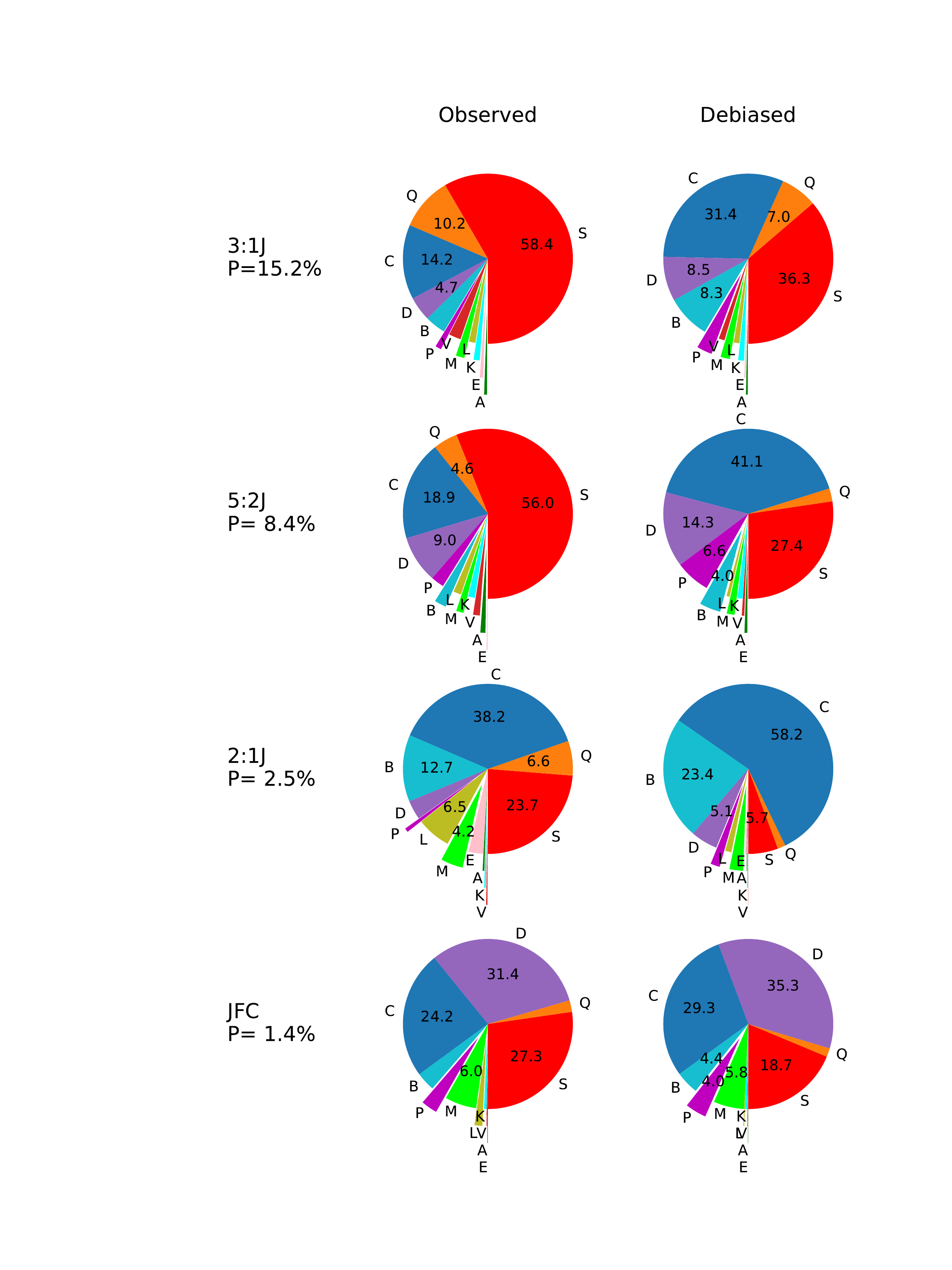}
\caption{Observed versus debiased compositional distribution of the MITHNEOS survey for the overall NEO population (top left), and individual escape/source regions. Regions are ordered by increasing distance from the Sun. Each percentage indicates the average probability for an NEO in our dataset of originating from the source region. 11 objects in our dataset were not included in this plot due to uncertain spectral classification (Section\,\ref{sec:taxons}). This figure shows the compositional mixing of the NEO population and, therefore, of the asteroid belt at small object sizes, as well as the global trend of increasing dark-to-bright object ratio as a function of ER distance from the Sun.}
\label{fig:pie_charts_all}
\end{figure}

The observed and debiased compositional distributions of the overall NEO population and individual ERs are shown in Fig.~\ref{fig:pie_charts_all}. 
In agreement with previous works (e.g., \citealt{Tholen:1984, Zellner:1985, Tedesco:1987, Veeder:1989, Binzel:2004, Binzel:2019, Carry:2016, Perna:2018, Devogele:2019}), the overall observed population is dominated by silicate-rich (S- and Q-type) bodies, accounting for $\sim$66\% of objects in our dataset. 
The two most proficient ERs, the $\upnu_6$ and 3:1J complexes, exhibit very similar distributions with one another (Pearson correlation coefficient of 99.8\%). 
This is unsurprising considering that both ERs are dominantly fed by the inner belt population of asteroids. 
As we move towards ERs located at larger heliocentric distances, the observed fraction of silicate-rich objects gradually decreases and the fraction of carbonaceous objects (B-, C-, P-, D-type) increases, with the 2:1J and JFCs exhibiting the highest fractions (55--62\%) of carbonaceous objects. 
The relatively high fraction of silicate-rich objects among JFCs is here likely an artifact due to the low number of comets in our sample, as well as the correlated ER probabilities of the 3:1J, 5:2J and JFCs. 


As expected, bias-correction leads to greater fractional abundances of dark objects and smaller fractional abundances of bright objects in the NEO population. 
The abundance of silicate-rich objects decreases from 66\% to 42\% globally, while the abundance of carbonaceous objects increases from 23\% to 48\%. 
It is evident from Fig.~\ref{fig:pie_charts_all} that the effects of debiasing are both albedo-dependent and ER-dependent. 
The greater the difference in albedo between a specific taxonomic class of objects and the average population, the greater the change in fractional abundance. 
Similarly, the steeper the HFD of a given ER, the greater the change in dark-to-bright ratio when debiased. 
For instance, Hungaria NEOs are the least affected population owing to their flat HFD below H=19 ($\beta$=0.074), where most NEOs with high Hungaria ER probability are found in our dataset. 

Like the observed population, the debiased NEO population reflects a composition gradient in the main belt, with a decreasing fraction of silicate-rich bodies and an increasing fraction of carbonaceous bodies as a function of increasing heliocentric distance of the ER. 
In addition, NEOs from the various ERs exhibit a wide diversity of taxons supporting the idea of a well-mixed population of small (D$<$2\,km) asteroids in the main belt. 
This mixed distribution agrees well with the distribution of the smallest asteroids observed in the main belt (see following Section\,\ref{sec:NEOs_vs_MBAs}), and contrasts with the more radially stratified distribution of large (D$>$20\,km) asteroids \citep{Demeo:2013}. 
This difference between small and large asteroids is commonly attributed to the increased efficiency of the Yarkovsky drift, as well as dynamical pushes due to planetary encounters and resonant orbits toward smaller body sizes.


\subsection{Comparison with the asteroid belt}
\label{sec:NEOs_vs_MBAs}

\subsubsection{Global compositional match}
\label{sec:match}

\begin{figure}[h!]
\centering
\includegraphics[angle=0, width=0.45\linewidth, trim=0cm 0cm 0cm 0cm, clip]{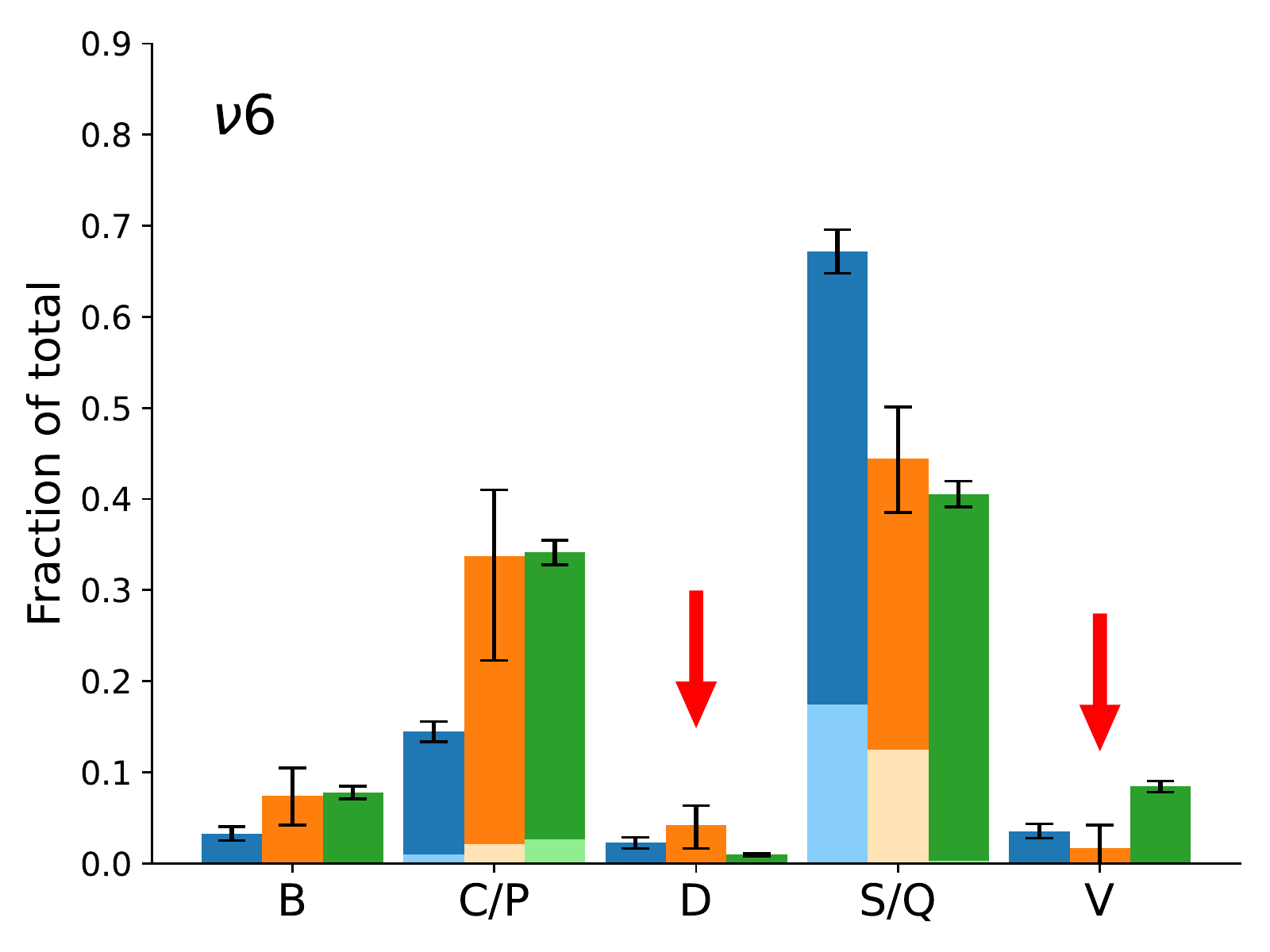}
\includegraphics[angle=0, width=0.45\linewidth, trim=0cm 0cm 0cm 0cm, clip]{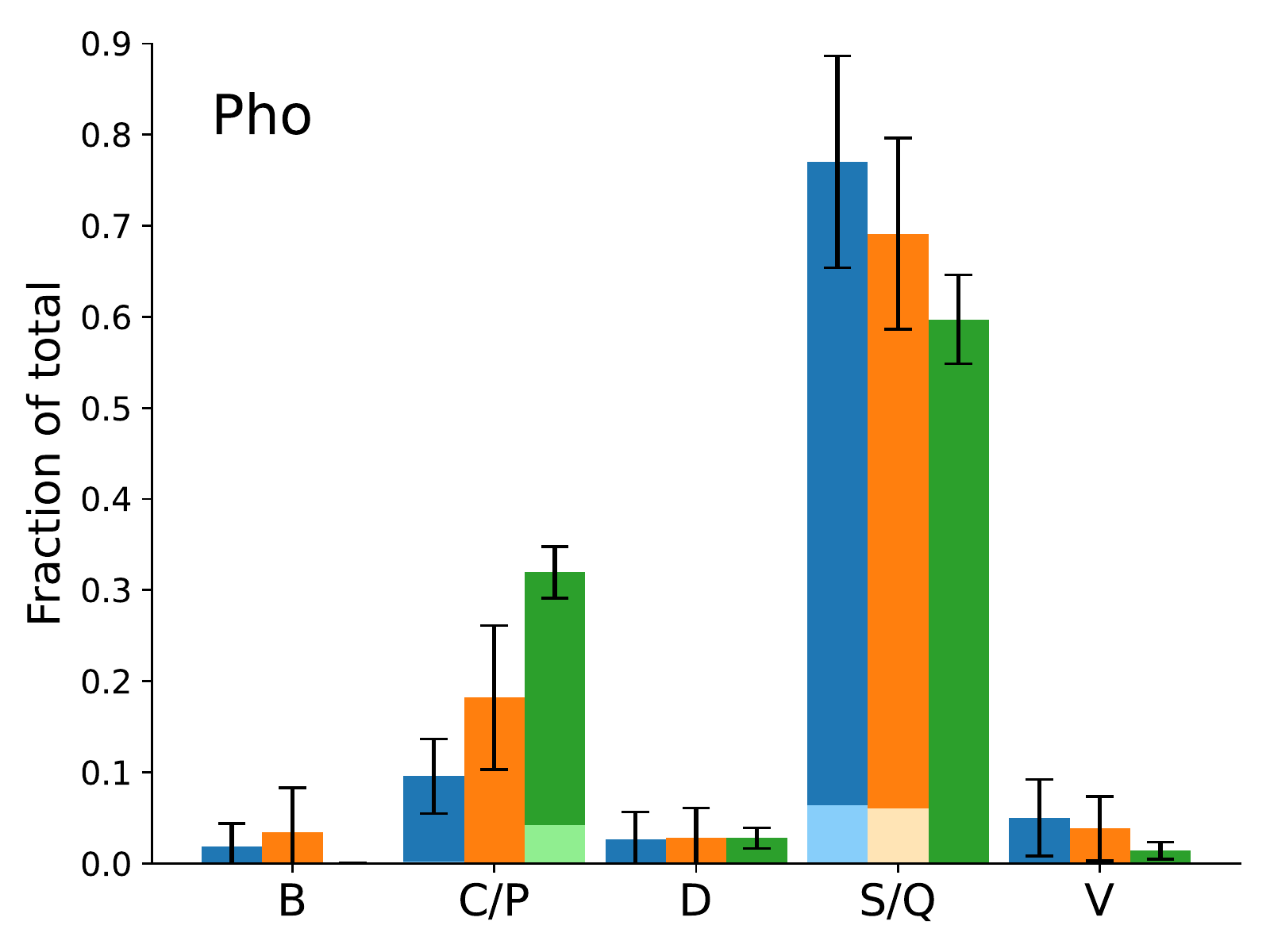}
\includegraphics[angle=0, width=0.45\linewidth, trim=0cm 0cm 0cm 0cm, clip]{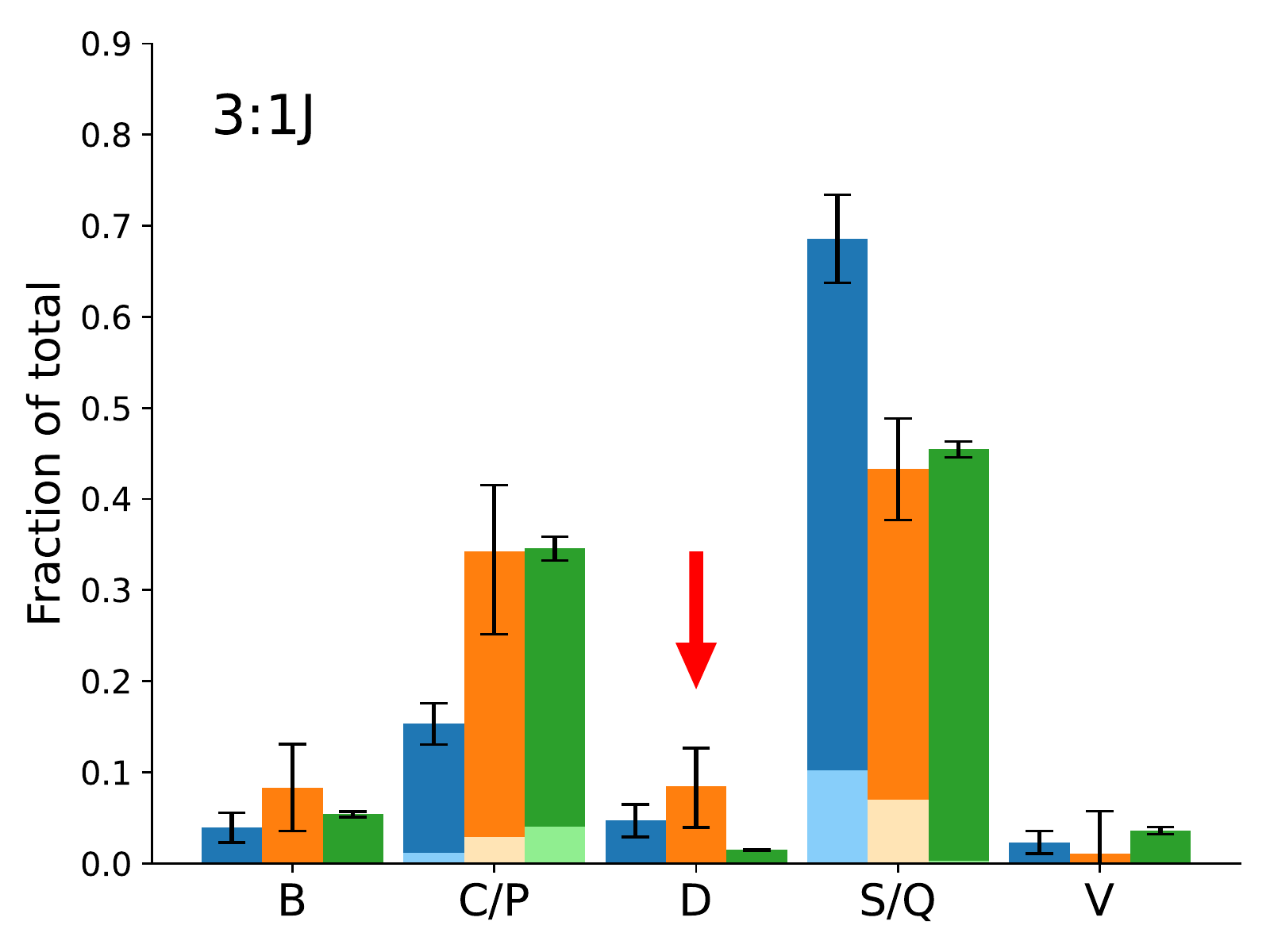}
\includegraphics[angle=0, width=0.45\linewidth, trim=0cm 0cm 0cm 0cm, clip]{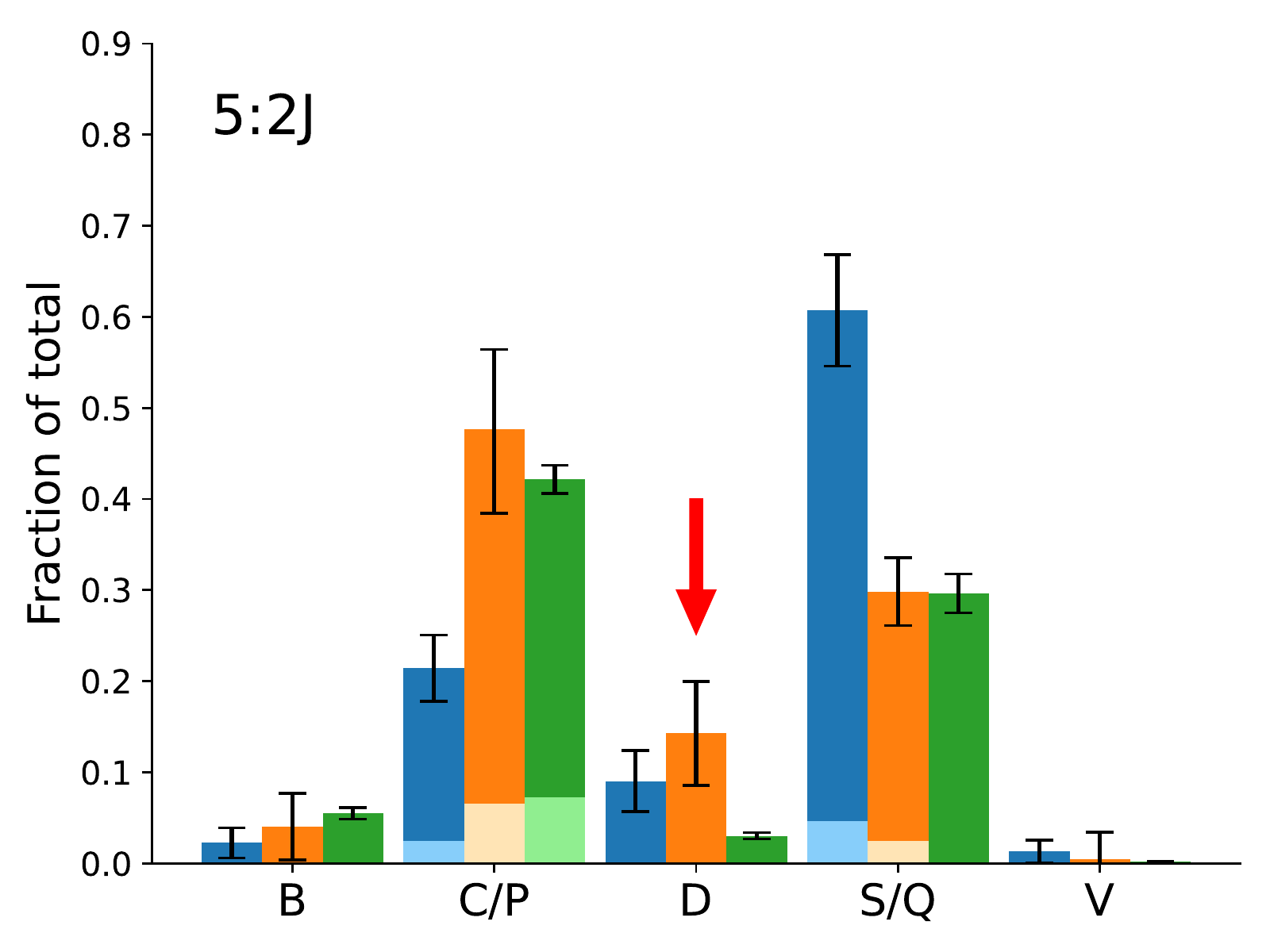}
\includegraphics[angle=0, width=0.45\linewidth, trim=0cm 0cm 0cm 0cm, clip]{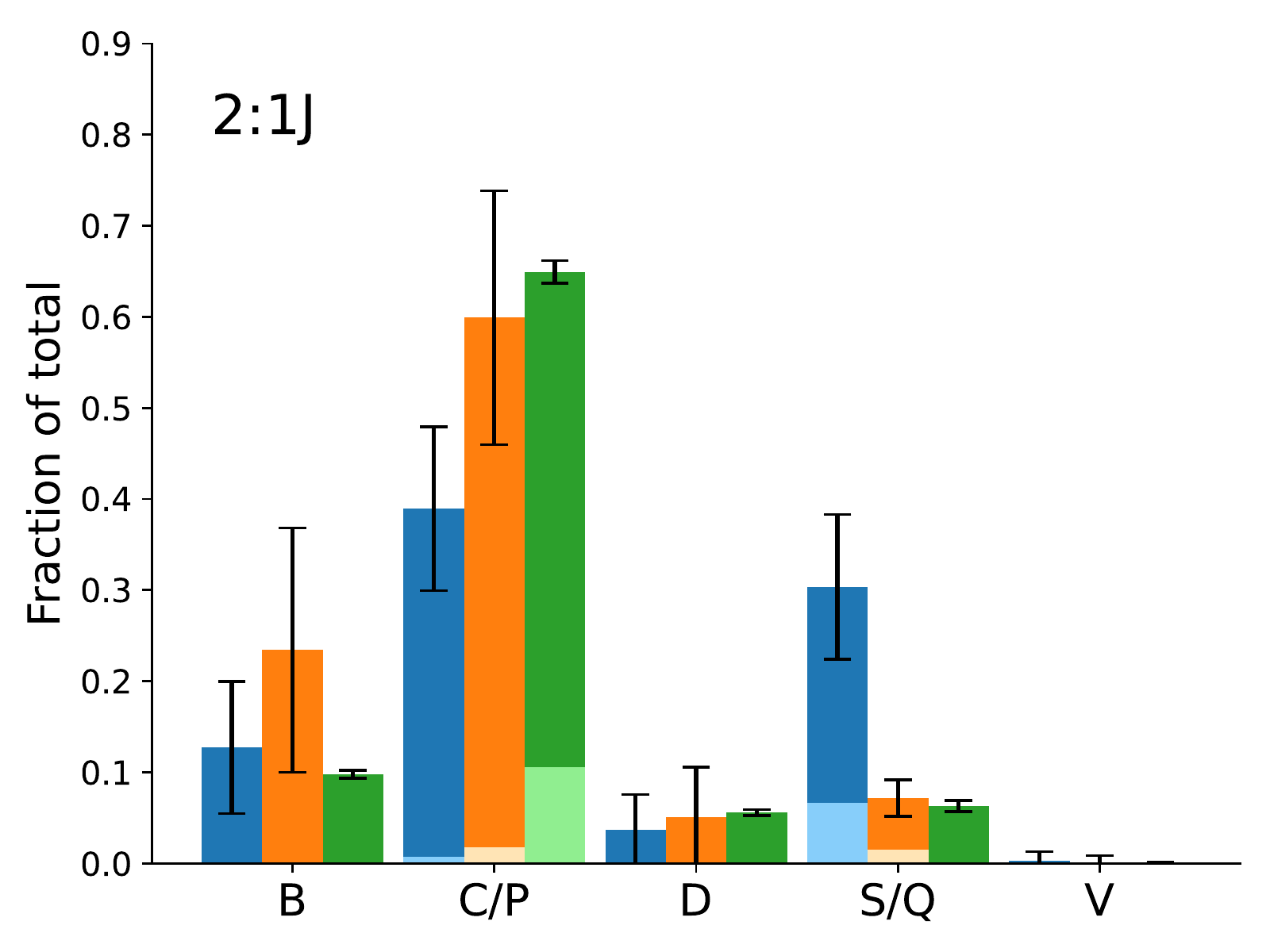}
\includegraphics[angle=0, width=0.45\linewidth, trim=0cm 0cm 0cm 0cm, clip]{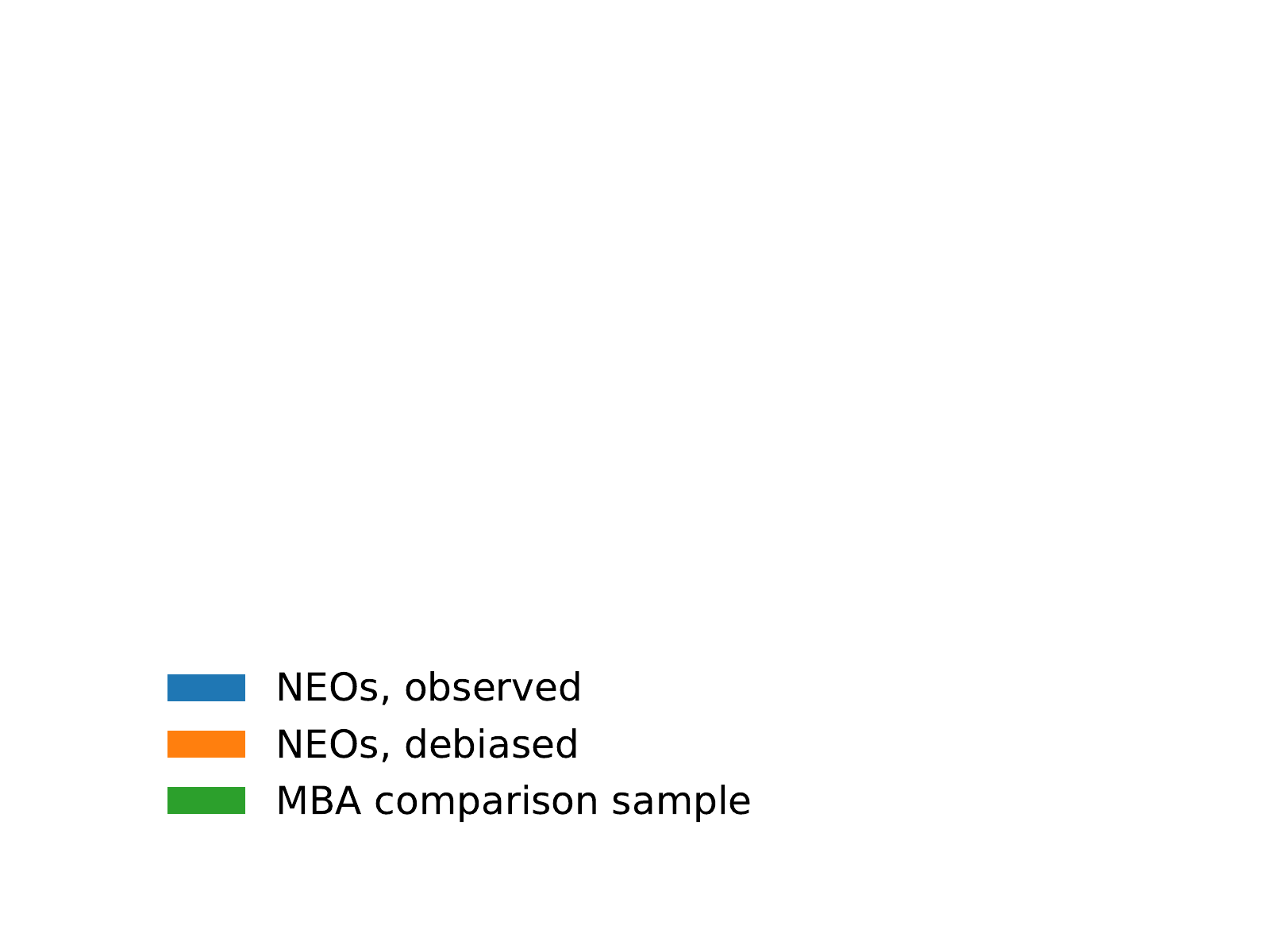}
 \caption{Observed (blue) and debiased (orange) fractional abundances of spectral classes of NEOs, compared to their predicted source populations in the asteroid belt (green). Escape regions are ordered by increasing distance from the Sun and their name is indicated in the upper left corner of the panels. See Section~\ref{sec:criteria} for a description of the selection criteria of the main-belt datasets. Carbonaceous C- and P-type objects were merged into a single class "C/P", and silicate-rich S and Q-type into "S/Q"  (see Section\,\ref{sec:classes}). For these two broad classes, the plotted bars are divided into two shades of color, with the darker shade corresponding to C- and S-types, and the lighter shade to P- and Q-types. 
 The least populated taxonomic classes (A-, K-, L-, E-, and M-types) are not displayed here for better readability. 
 A second version of the figure including every taxonomic classes is provided in Appendix\,\ref{sec:app_D}. 
 This figure highlights the compositional match between NEOs and their predicted source regions after bias correction. The red arrows underline local composition differences discussed in Section\,\ref{sec:diff}.}
\label{fig:debAbundances_MBAalb}
\end{figure}

Fig.~\ref{fig:debAbundances_MBAalb} compares the observed and debiased compositional distributions of NEOs derived in this work with their corresponding source populations in the asteroid belt, as defined in Table\,\ref{tab:ERs}. 
For readability purpose, only the main (most populated) taxonomic classes of asteroids (B-, C/P-, D-, S/Q-, and V-types) are shown. 
A second version of the figure showing every classes is provided in Appendix\,\ref{sec:app_D}. 
Due to the low number of Hungaria MBAs in the relevant range of sizes (Section\,\ref{sec:criteria}), this population of objects is not shown in the figure. 

The observed fractional abundances of each taxonomic complex were assigned Poisson error bars ($\sigma_N = \sqrt{N}$). 
The error bars of the debiased fractional abundances were derived by propagating the Poisson uncertainty on the observed fractional abundances, the uncertainty on the albedo and the uncertainty on $\beta$ as computed by \citet{Granvik:2018} for 17$<$H$<$22 (see their Table 4). 
Taxonomic abundances for the debiased NEOs and the comparison MBA source populations are provided in Table~\ref{tab:abundances}. 

Globally, the debiased taxonomic abundances of NEOs closely match the taxonomic distribution of their predicted source populations in the asteroid belt. 
In Table~\ref{tab:stats}, we provide the Pearson $r$ coefficient and corresponding $p$-value for a correlation between the NEO and MBA populations. 
For each ER, we also provide the maximum difference between the measured taxonomic ratios of the NEO and MBA populations. 
In every cases, the bias correction greatly improves the correlation between the NEOs and their source populations, and significantly decreases the compositional gap between these populations. 

The global compositional match between NEOs and their predicted source populations in the asteroid belt provides a direct validation of the NEO migration model of \citet{Granvik:2018} used in this work. 
This compositional match also argues against any drastic composition size trend in the asteroid belt, in the range of diameters studied here (i.e., from $\sim$10--4\,km for the MBA dataset, down to typical NEO sizes, $\sim$3\,km--100\,m). 
Deviations from this compositional agreement are discussed in following Section\,\ref{sec:diff}. 

\begin{deluxetable}{cc|rrrrr}[h!]
\label{tab:abundances}
\tabletypesize{ \scriptsize}
\tablecaption{Taxonomic ratios of the observed and debiased NEO populations, compared to their source regions in the asteroid belt.}
\tablehead{Source & Taxon &  \multicolumn{1}{c}{\% NEOs} &  \multicolumn{1}{c}{\% NEOs} & 
\multicolumn{1}{c}{\% MBAs} &
\multicolumn{1}{c}{difference} &
\multicolumn{1}{c}{difference} \\
region & & (observed) & (debiased) & & (observed) & (debiased) }
\startdata
$\upnu_6$ &      A & 0.5$\pm$0.3 & 0.3$\pm$1.9  & 0.2$\pm$0.1 & 0.3$\pm$0.3 & 0.1$\pm$1.9 \\
          &      B & 3.3$\pm$0.8 & $7.4^{+3.1}_{-3.2}$  & 7.8$\pm$0.7 & 4.5$\pm$1.0 & $0.4^{+3.1}_{-3.3}$ \\
          &    C/P & 14.5$\pm$1.1 & $33.7^{+7.3}_{-11.5}$  & 34.1$\pm$1.4 & 19.7$\pm$1.8 & $0.4^{+7.4}_{-11.5}$ \\
          &      D & 2.2$\pm$0.6 & $4.2^{+2.1}_{-2.5}$  & 0.9$\pm$0.1 & 1.3$\pm$0.6 & $3.3^{+2.1}_{-2.6}$ \\
          &    S/Q & 67.2$\pm$2.4 & $44.4^{+5.7}_{-5.9}$  & 40.5$\pm$1.4 & 26.7$\pm$2.8 & $3.9^{+5.9}_{-6.1}$ \\
          &      K & 1.0$\pm$0.4 & 1.0$\pm$0.8  & 1.9$\pm$0.1 & 0.9$\pm$0.4 & 0.9$\pm$0.8 \\
          &      L & 3.8$\pm$0.8 & 3.7$\pm$1.6  & 2.7$\pm$0.2 & 1.2$\pm$0.8 & 1.0$\pm$1.6 \\
          &      V & 3.5$\pm$0.8 & 1.7$\pm$2.5  & 8.4$\pm$0.6 & 4.9$\pm$1.0 & 6.8$\pm$2.6 \\
          &      E & 1.2$\pm$0.4 & 0.4$\pm$3.6  & 0.2$\pm$0.0 & 0.9$\pm$0.4 & 0.2$\pm$3.6 \\
          &      M & 2.9$\pm$0.7 & 3.3$\pm$1.2  & 3.2$\pm$0.0 & 0.4$\pm$0.7 & 0.0$\pm$1.2 \\
Pho &      A & 0.4$\pm$1.2 & 0.4$\pm$1.6  & 0.0$\pm$0.0 & 0.4$\pm$1.2 & 0.4$\pm$1.6 \\
    &      B & 1.8$\pm$2.5 & 3.4$\pm$4.9  & 0.0$\pm$0.0 & 1.8$\pm$2.5 & 3.4$\pm$4.9 \\
    &    C/P & 9.6$\pm$4.1 & 18.2$\pm$7.9  & 31.9$\pm$2.8 & 22.4$\pm$5.0 & 13.7$\pm$8.4 \\
    &      D & 2.6$\pm$3.0 & 2.8$\pm$3.3  & 2.8$\pm$1.1 & 0.2$\pm$3.2 & 0.0$\pm$3.5 \\
    &    S/Q & 77.0$\pm$11.6 & 69.1$\pm$10.5  & 59.7$\pm$4.9 & 17.3$\pm$12.6 & 9.4$\pm$11.6 \\
    &      K & 1.5$\pm$2.3 & 0.4$\pm$0.8  & 1.4$\pm$0.9 & 0.2$\pm$2.5 & 1.0$\pm$1.2 \\
    &      L & 1.0$\pm$1.8 & 0.8$\pm$1.5  & 0.0$\pm$0.0 & 1.0$\pm$1.8 & 0.8$\pm$1.5 \\
    &      V & 5.0$\pm$4.2 & 3.8$\pm$3.5  & 1.4$\pm$0.9 & 3.6$\pm$4.3 & 2.4$\pm$3.6 \\
    &      E & 0.4$\pm$1.1 & 0.2$\pm$2.1  & 0.0$\pm$0.0 & 0.4$\pm$1.1 & 0.2$\pm$2.1 \\
    &      M & 0.6$\pm$1.5 & 0.8$\pm$1.9  & 2.8$\pm$0.2 & 2.1$\pm$1.5 & 2.0$\pm$1.9 \\
3:1J &      A & 0.5$\pm$0.6 & 0.3$\pm$3.4  & 0.2$\pm$0.0 & 0.3$\pm$0.6 & 0.1$\pm$3.4 \\
     &      B & 3.9$\pm$1.6 & $8.3^{+4.7}_{-4.8}$  & 5.4$\pm$0.3 & 1.5$\pm$1.7 & $2.9^{+4.7}_{-4.8}$ \\
     &    C/P & 15.3$\pm$2.3 & $34.3^{+7.3}_{-9.1}$  & 34.6$\pm$1.3 & 19.3$\pm$2.6 & $0.3^{+7.4}_{-9.2}$ \\
     &      D & 4.7$\pm$1.8 & $8.5^{+4.2}_{-4.5}$  & 1.5$\pm$0.0 & 3.2$\pm$1.8 & $7.0^{+4.2}_{-4.5}$ \\
     &    S/Q & 68.6$\pm$4.8 & 43.3$\pm$5.6  & 45.4$\pm$0.9 & 23.1$\pm$4.9 & 2.2$\pm$5.6 \\
     &      K & 1.2$\pm$0.9 & 1.1$\pm$1.4  & 2.3$\pm$0.1 & 1.1$\pm$0.9 & 1.2$\pm$1.4 \\
     &      L & 1.2$\pm$0.9 & 1.1$\pm$1.6  & 4.1$\pm$0.2 & 2.8$\pm$0.9 & 3.0$\pm$1.6 \\
     &      V & 2.3$\pm$1.2 & 1.1$\pm$4.6  & 3.6$\pm$0.4 & 1.3$\pm$1.3 & 2.5$\pm$4.7 \\
     &      E & 0.6$\pm$0.7 & 0.2$\pm$6.5  & 0.1$\pm$0.0 & 0.5$\pm$0.7 & 0.1$\pm$6.5 \\
     &      M & 1.6$\pm$1.1 & 1.8$\pm$1.2  & 2.8$\pm$0.0 & 1.2$\pm$1.1 & 1.0$\pm$1.2 \\
5:2J &      A & 1.0$\pm$1.1 & 0.5$\pm$2.1  & 0.2$\pm$0.1 & 0.8$\pm$1.1 & 0.4$\pm$2.1 \\
     &      B & 2.2$\pm$1.7 & 4.0$\pm$3.7  & 5.5$\pm$0.6 & 3.2$\pm$1.8 & 1.4$\pm$3.7 \\
     &    C/P & 21.4$\pm$3.6 & $47.7^{+8.7}_{-9.2}$  & 42.1$\pm$1.5 & 20.7$\pm$4.0 & $5.5^{+8.9}_{-9.4}$ \\
     &      D & 9.0$\pm$3.3 & $14.3^{+5.7}_{-5.8}$  & 3.0$\pm$0.4 & 6.0$\pm$3.4 & $11.3^{+5.7}_{-5.8}$ \\
     &    S/Q & 60.7$\pm$6.1 & 29.8$\pm$3.7  & 29.6$\pm$2.1 & 31.1$\pm$6.5 & 0.2$\pm$4.3 \\
     &      K & 1.3$\pm$1.3 & 1.0$\pm$1.1  & 8.8$\pm$0.3 & 7.5$\pm$1.3 & 7.8$\pm$1.1 \\
     &      L & 1.5$\pm$1.3 & 0.6$\pm$1.3  & 4.3$\pm$0.4 & 2.8$\pm$1.4 & 3.7$\pm$1.4 \\
     &      V & 1.3$\pm$1.3 & 0.5$\pm$2.9  & 0.2$\pm$0.0 & 1.1$\pm$1.3 & 0.3$\pm$2.9 \\
     &      E & 0.2$\pm$0.5 & 0.1$\pm$4.8  & 0.1$\pm$0.0 & 0.1$\pm$0.5 & 0.0$\pm$4.8 \\
     &      M & 1.4$\pm$1.3 & 1.5$\pm$1.4  & 6.2$\pm$0.0 & 4.8$\pm$1.3 & 4.7$\pm$1.4 \\
2:1J &      A & 0.3$\pm$1.1 & 0.1$\pm$0.6  & 0.0$\pm$0.0 & 0.3$\pm$1.1 & 0.1$\pm$0.6 \\
     &      B & 12.7$\pm$7.3 & 23.4$\pm$13.4  & 9.8$\pm$0.4 & 2.9$\pm$7.3 & 13.6$\pm$13.4 \\
     &    C/P & 38.9$\pm$9.0 & $59.9^{+13.9}_{-14.0}$  & 64.9$\pm$1.3 & 26.0$\pm$9.1 & 5.0$\pm$14.0 \\
     &      D & 3.6$\pm$3.9 & 5.1$\pm$5.5  & 5.6$\pm$0.3 & 1.9$\pm$3.9 & 0.5$\pm$5.5 \\
     &    S/Q & 30.3$\pm$7.9 & 7.2$\pm$2.0  & 6.3$\pm$0.6 & 24.1$\pm$8.0 & 0.9$\pm$2.1 \\
     &      K & 0.3$\pm$1.1 & 0.1$\pm$0.6  & 5.8$\pm$0.6 & 5.5$\pm$1.2 & 5.7$\pm$0.9 \\
     &      L & 6.5$\pm$5.2 & 1.2$\pm$1.0  & 2.4$\pm$0.2 & 4.1$\pm$5.2 & 1.2$\pm$1.1 \\
     &      V & 0.2$\pm$1.0 & 0.0$\pm$0.8  & 0.1$\pm$0.0 & 0.2$\pm$1.0 & 0.0$\pm$0.8 \\
     &      E & 2.8$\pm$3.4 & 0.3$\pm$1.5  & 0.1$\pm$0.0 & 2.7$\pm$3.4 & 0.3$\pm$1.5 \\
     &      M & 4.2$\pm$4.2 & 2.6$\pm$2.6  & 5.1$\pm$0.0 & 0.9$\pm$4.2 & 2.5$\pm$2.6 \\
\enddata
\end{deluxetable}

\begin{deluxetable}{c|ccrc|ccrc}[h!]
\tabletypesize{ \scriptsize}
\tablecaption{Results of statistic tests used to assess the compositional similarities of the observed and debiased compositional distributions of NEOs, compared to their source regions in the asteroid belt.}
\tablehead{Source region & \multicolumn{3}{c}{Observed NEOs vs. MBAs} & \multicolumn{3}{c}{Debiased NEOs vs. MBAs} \\
& Coeff. of correlation$^*$ & $p$-value$^*$ & max. diff.$^{**}$ &  Taxon & Coeff. of correlation$^*$ & $p$-value$^*$ & max. diff.$^{**}$ & Taxon }
\startdata
$\upnu_6$ & 84.4\% & 2.2e-03 & 26.7$\pm$2.8\% & S/Q 	& 98.4\% & 2.7e-07 & 6.8$\pm$2.6\% & V \\
Pho & 91.8\% & 1.8e-04 & 22.4$\pm$5.0\% & C/P 	& 96.4\% & 7.4e-06 & 13.7$\pm$8.4\% & C/P \\
3:1J & 88.3\% & 7.0e-04 & 23.1$\pm$4.9\% & S/Q 	& 98.3\% & 3.8e-07 & $7.0^{+4.2}_{-4.5}$\% & D \\
5:2J & 72.7\% & 1.7e-02 & 31.1$\pm$6.5\% & S/Q 	& 94.7\% & 3.2e-05 & $11.3^{+5.7}_{-5.8}$\% & D \\
2:1J & 79.0\% & 6.5e-03 & 26.0$\pm$9.0\% & C/P 	& 96.3\% & 8.0e-06 & 13.6$\pm$13.4\% & B \\
\enddata
\tablenotetext{}{$^*$Coefficient of correlation and $p$-value from Pearson $r$ test.\\ 
$^{**}$Maximum difference in fractional abundance between the populations, with corresponding taxonomic class indicated in the following column.}
\label{tab:stats}
\end{deluxetable}

\subsubsection{Local compositional differences}
\label{sec:diff}

Deviations from the global compositional match between NEOs and their predicted ERs might provide insights to the mechanisms responsible for delivering main-belt material to the NEO space. 
For instance, the excess of a given taxonomic type of NEOs compared to their source population can indicate the presence of a collisional family of small asteroids near the ER, below the size detection limit of current surveys of the main belt. 

In this Section, we list and discuss possible origins for the compositional differences found in our dataset. 
We refer to an ``excess" (or ``lack") of taxonomic class when the fractional abundance of that class in the NEO population exceeds (is less than) its abundance in the source population. 
We deliberately ignore the rare K and L classes, which are often indistinguishable from other spectral classes in visible wavelengths \citep{Demeo:2013}.

\vspace{2mm}
\paragraph{\centering Silicate-rich surface evolution\\}

In agreement with previous works (e.g., \citealt{Binzel:2010}), we find that the NEO population comprises a larger abundance of Q-type bodies among silicate-rich bodies compared to the asteroid belt (Fig.\,\ref{fig:debAbundances_MBAalb}). 
For instance, $\upnu_6$ NEOs have Q/(Q+S)=26\% in our dataset, whereas $\upnu_6$ MBAs have Q/(Q+S)=0.7\%. 
Q-type asteroids are commonly interpreted to have fresh ordinary-chondrite-like surface compositions that have been recently refreshed and not altered by space weathering processes, whereas S-types would be ordinary chondrite asteroids whose surfaces have experienced increasing amounts of weathering as a result of being exposed to the space environment for longer times. Several mechanisms -- including planetary encounters, Yarkovsky-O'Keefe-Radzievskii-Paddack (YORP) spin up and thermal cracking -- have been proposed to be responsible for refreshing asteroid surfaces, preferentially in the NEO space (\citealt{Marchi:2006, Binzel:2010, Nesvorny:2010, Delbo:2014, DeMeo:2014_Mars, Polishook:2014, Graves:2018, Graves:2019}). 
Q-to-S ratios derived from SDSS, however, should be interpreted with caution because Q and S-type spectra are hardly distinguishable at low spectral resolution in visible wavelengths. 

\vspace{2mm}
\paragraph{\centering Lack of V-type NEOs from the inner belt\\}

Most V-type asteroids and basaltic meteorites found on Earth originated from the Vesta family, although alternative sources for these objects have also been proposed \citep{Moskovitz:2008, Solontoi:2012, Leith:2017, Migliorini:2021}. 
V-type asteroids only account for 1.7$\pm$2.5\% of $\upnu_6$ NEOs in our dataset, while representing 8.4$\pm$0.6\% of the inner-belt asteroids. 
This discrepancy cannot be explained as misidentification of V-type bodies in SDSS: 
using optical and near-infrared spectroscopy, \citet{Moskovitz:2008} confirmed the spectral type of 10 out of 11 V-type asteroids (91\%) identified as V-type candidates based on photometric properties in SDSS. Similarly, \citet{Migliorini:2021} confirmed the basaltic nature of 16 out of 23 V-type asteroids (70\%) selected from SDSS. 
One can therefore reasonable assume that around 80\% of asteroids identified as V-type by \citet{Demeo:2013} are real V-types, still implying that V-type are $\sim$4 times less abundant among NEOs compared to the inner belt. 



Alternatively, the lack of V-type NEOs may stem from the central location of the Vesta family in the inner belt, away from any strong resonance with the giant planets. 
Less likely explanation is the old age of the Vesta family ($\sim$1\,Gy; \citealt{Marzari:1999, Carruba:2005, Nesvorny:2008,  Marchi:2012, Schenk:2012, Nesvorny:2015_families, Spoto:2015, Carruba:2016}), 
considering that Flora, an even older inner-belt family ($>$1\,Gy; \citealt{Broz:2013, Carruba:2016}) located close to the $\upnu_6$ secular resonance, was identified as the main source of S-type NEOs \citep{Vernazza:2008}. 
It should be noted, however, that it takes on average only 80\,Myr for a 100-m Vestoid (a Vesta family member) to travel from Vesta to q=1.3\,au \citep{Unsalan:2019}. 
Therefore, the argument of the central location of Vesta in the asteroid belt does not appeal as a fully satisfactory answer either. 
Future modelling work should investigate further the fractional contribution of collisional families to NEOs as a function of family size, age and location in the asteroid belt.

\vspace{2mm}
\paragraph{\centering Excess of D-type NEOs from the inner and middle belt\\} 

We further note a $\sim$2-$\upsigma$ excess of D-type NEOs coming from the 5:2J ER. 
This excess is also seen in the $\upnu_6$ and 3:1J ERs at the $\sim$1.5-$\upsigma$ statistical level. 
While the statistical significance of this excess might seem minor, the fact that it is observed in three distinct ERs provides credibility to its existence. 
In addition, we remind that the NEO taxonomic ratios reported here result from bias-correcting the sample using average albedo values measured from the MBA dataset (i.e., an average albedo of $\sim$8\% for the D-types; Section\,\ref{sec:albedos}). Using NEO values from Table\,\ref{tab:albedos} (i.e., 4\% for the D-types) amplifies the gap between the fractional abundances of D-type NEOs and their source populations: 
this finding appears to be a robust result. 

D-type NEOs being more abundant than their larger counterparts in the main belt may be indicative of a global trend of increasing fractions of D-type bodies as size decreases. 
This would be consistent with the finding of \citet{Perna:2018} that D-types are fractionally more abundant among small-size (D$<$500\,m) NEOs compared to larger sizes. 
It should be noted, however, that other spectroscopic surveys of small NEOs did not confirm this trend \citep{Devogele:2019}. 

As discussed by \citet{Perna:2018}, the existence of such a trend would hold important implications for our understanding of the early delivery of water and prebiotic material to the primitive Earth. 
Indeed, the idea that comets constituted the main suppliers of these elements was recently revised owing to their higher deuterium-to-hydrogen ratios with respect to ocean water \citep{Bockelee:2015,Altwegg:2015}. 
As a consequence, it is now commonly admitted that another population of objects must have dominated the early flux of water and organics to the Earth.   
Carbonaceous asteroids and more specifically D-type bodies are prime candidates in that regard, being among the most compositionally pristine bodies in the Solar System 
\citep{Hiroi:2001}. 
As such, increasing fractions of D-type bodies at smaller sizes may imply that these objects contributed more to the early delivery of the prebiotic material to the Earth than previously estimated. 


Several effects may account for the observed excess of D-type bodies in the NEO population: 
First, a fraction of C- and P-type carbonaceous bodies may preferentially evolve into D-type spectrum towards smaller sizes and closer distances to the Sun. 
This was evidenced by laboratory experiments simulating the effects of space weathering on carbonaceous chondrite meteorites, finding that fresh D-type surfaces could be transformed into C- and P-types due to solar wind \citep{Lantz:2018}. 
In addition, \citet{Hasegawa:2022} found out that asteroid (596)~Scheila, which was resurfaced by an impact in 2010, experienced a significant change of its NIR spectral slope, from T-type (that is, intermediate between P- and D-) to D-type, as a consequence of the impact. 
A genetic link between C-, P- and D-type asteroids is further supported by the similarities of these bodies in bulk densities, suggesting they share a common bulk composition \citep{Vernazza:2021}. 
Taken together, these studies suggest that D-type asteroids  may represent the fresh, unweathered version of C- and P-type bodies. 
In that context, the overabundance of D-type NEOs may be explained by the fact that some of these objects come from the population of C- and P-type asteroids in the main belt, but they exhibit younger surfaces on average owing to their shorter collisional lifetimes. 
As discussed by \citet{Vernazza:2021} and \citet{Hasegawa:2022}, an important implication for a direct link between C-, P- and D-type asteroids would be that a much larger fraction of the main belt than previously estimated could be implanted objects from the transplanetary and/or transneptunian region \citep{Levison:2009}.

Second, the excess of D-type NEOs may arise from an increasing fraction of D-type bodies at smaller sizes {\it within} the asteroid belt. This increase could come from the existence of a family of small D-type asteroids, with sizes below the characterization limit of current main-belt surveys, formed by the fragmentation of a larger D-type body.
Considering that the excess is most obvious in the case of the 5:2J and 3:1J MMRs, one should expect the population or family to be located in the middle belt between these two resonances.

Third, as proposed by \citet{Perna:2018}, D-type bodies may fragment more easily in NEO space compared to other spectral types of asteroids due to their higher fragility, either through collisions and/or thermal cracking \citep{Delbo:2014, Granvik:2016}. 
However, D-types bodies are also believed to be highly porous objects \citep{Vernazza:2015}, implying that compaction, not fragmentation, may preferentially occur in a collision \citep{Housen:1999, Housen:2018}. 
On the other hand, preferential fragmentation through thermal cracking  would be consistent with the dark surfaces of D-type asteroids: the lower the albedo, the more incident light an object absorbs, and the more it is prone to fragment due to strong thermal cycles.



Finally, it could be argued that the excess of D-type NEOs coming from the 3:1J and 5:2J ERs 
may also be an artifact coming from the probabilistic nature of our dynamical model. 
Indeed, NEOs coming from outer ERs (the 3:1J, 5:2J, 2:1J and JFCs) often exhibit correlated ER probabilities owing to their orbital similarities (an extreme example happens when bodies in chaotic resonances get their eccentricities and inclinations pumped up to cometary-like Jupiter Tisserand $T_J<3$; \citealt{Farinella:1994,Gladman:1997}). 
The 3:1J and 5:2J NEO samples therefore could be biased by D-type bodies from the JFCs. 
However, 
excluding objects with JFC probabilities larger than 50\% (5 objects) or even 10\% (32 objects) from our dataset has virtually no effect on the resulting fractional abundance of D-type NEOs from the 3:1J and 5:2J: a contamination bias appears unlikely in this case.

Whether a population of small D-type asteroids trully exists in the asteroid belt will be easily tested by future large photometric surveys such as the Vera Rubin Observatory's {\it Legacy Survey of Space and Time} (LSST), that will provide photometric measurements for several million asteroids down to a few hundred meters in size \citep{Jones:2009}. 

\section{Summary}
\label{sec:summary}

We presented 491 new NIR spectra of 420 NEOs collected mainly between January 2015 and February 2021 as part of the MITHNEOS survey, that we combined with earlier measurements \citep{Binzel:2019} to investigate the compositional distribution of the NEO population. 
This population was then divided into individual ERs, 
and bias-corrected to 
obtain their intrinsic compositional distribution. 
Our results can be summarized as follows:

\begin{itemize}
    \item[$\square$] The bias-corrected NEO distribution reflects well the overall composition gradient and radial mixing of the asteroid belt \citep{Demeo:2013,Demeo:2014}. 
Fractional abundances of the taxonomic classes correlate with the location of the ERs, with decreasing fractions of silicate-rich (S- and Q-types) bodies and increasing fractions of carbonaceous (B-, C-, D- and P-types) bodies as a function of the heliocentric distance of the ERs. 
\item[$\square$] The close compositional match between NEOs and their source populations in the asteroid belt validates numerical models used to predict NEO ERs in the Solar System \citep{Granvik:2018}, and argues against any strong composition change with size in the asteroid belt between $\sim$5\,km down to $\sim$100\,m, with the following two noteworthy exceptions: 
\item[$\square$] The paucity of V-type bodies coming from the $\upnu_6$ compared to their abundance in the inner belt may stem from the average central location of the Vesta family in the inner belt, away from any strong resonance with the giant planets. 
However, considering the short migration timescales of NEO-sized Vestoids, future modelling work would be needed to better understanding the contribution of collisional family to NEOs as a function of family size, age and location in the belt. 
\item[$\square$] The over-abundance of D-type NEOs coming from 5:2J and, to a lesser extend, the 3:1J and $\upnu_6$ may hint at the existence of a large population of small D-type bodies (possibly a collisional family) in the asteroid belt (probably the middle region). Alternatively, this excess could be due to the fact that NEOs exhibit, on average, younger surfaces than their larger main-belt counterparts, and that D-type surfaces represent fresher versions of C- and P-type surfaces \citep{Lantz:2018, Hasegawa:2022}. A third possibility is that D-type bodies fragment more often in the NEO space compared to other bodies \citep{Granvik:2016, Perna:2018}. 
Future large spectroscopic surveys like LSST will allow testing these different hypotheses.

\item[$\square$] No additional evidence for the existence of collisional families in the asteroid belt, below the detection limit of current surveys, was found in this work. 
\end{itemize}


\section*{Acknowledgements}

We thank David Tholen and Julia de Le\'on for helpful discussion, and the anonymous reviewer for insightful comments and suggestions. Observations reported here were obtained at the NASA Infrared Telescope Facility, which is operated by the University of Hawaii under under contract 80HQTR19D0030 with the National Aeronautics and Space Administration. The authors acknowledge the sacred nature of Maunakea, and appreciate the opportunity to observe from the mountain. The MIT component of this work is supported by NASA grant 80NSSC18K0849. Any opinions, findings, and conclusions or recommendations expressed in this article are those of the authors and do not necessarily reflect the views of the National Aeronautics and Space Administration.

\DIFdelend \bibliographystyle{aasjournal}
\bibliography{references}

\appendix
\section{New near-infrared spectral data of NEOs}
\label{sec:thumbnails}

\begin{figure}[h!]
\centering
\includegraphics[angle=0, width=\linewidth, trim=0cm 0cm 0cm 0cm, clip]{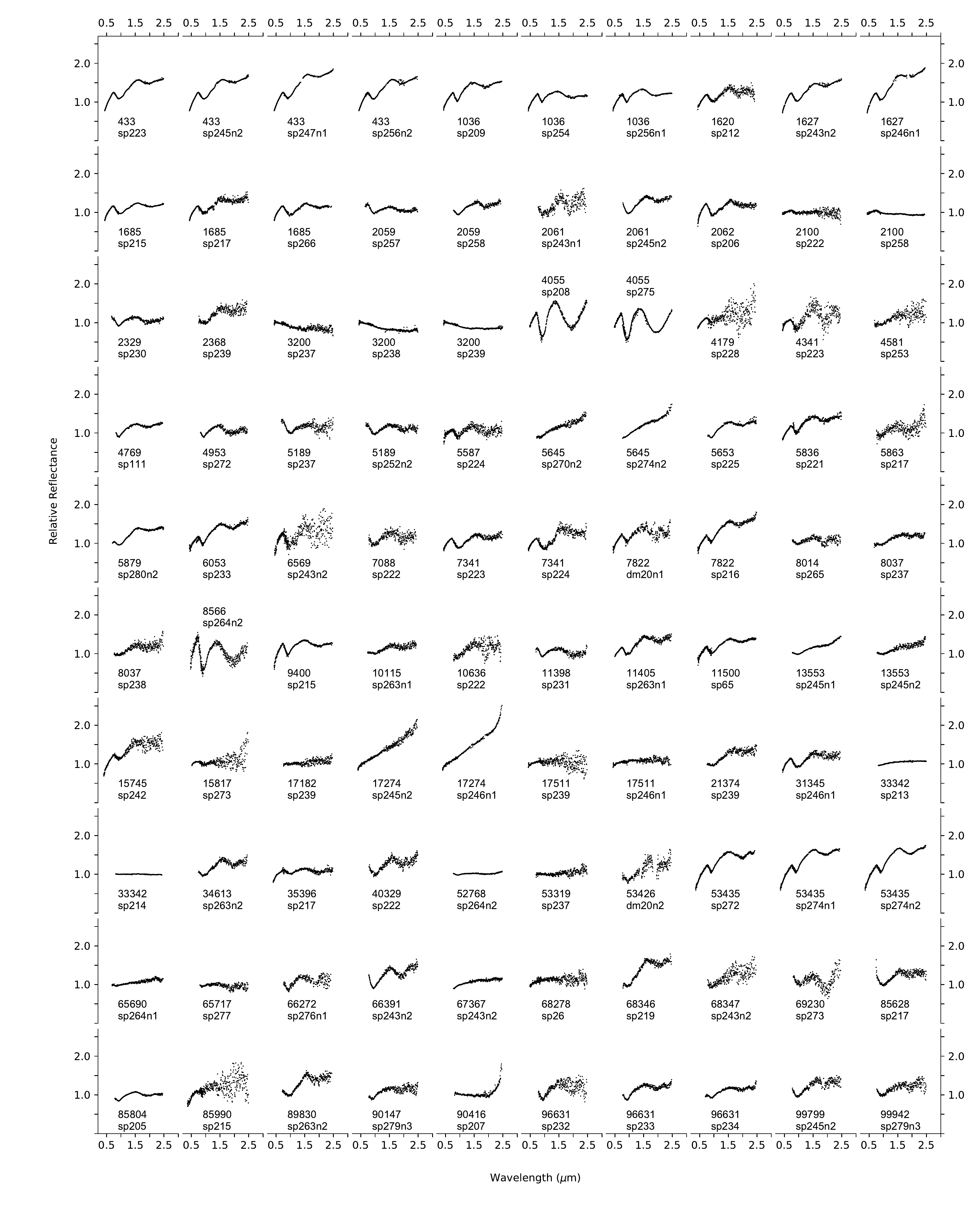}
\caption{Near-infrared spectral data of NEOs acquired mainly between January 2015 and February 2021 as part of the MITHNEOS program. The number and/or designation of each asteroid and the name of the observing run are provided below the spectra. The complete figure set (5 figures) is available in the online journal.}
\label{fig:thumbnails1}
\end{figure}

We present thumbnail versions of our reported spectral results in Fig.\,\ref{fig:thumbnails1}. 
The name of the asteroid and observing semester are indicated below each spectrum. 
Corresponding observing conditions are provided in Table~\ref{tab:suppmat} in Appendix~\ref{sec:app_E}.








\newpage

\section{Thermal parameters}
\label{sec:app_C}

We provide the thermal parameters obtained by NEATM modelling \citep{Harris:1998} and used to fit and remove the thermal tails in the asteroid spectra in Table\,\ref{tab:thermal}. 
These parameters are provided to allow reproducibility of the modelling work by other researchers. However, we emphasize that the values should be interpreted with care owing to the large uncertainties inherent to the modelling method (e.g., \citealt{Nugent:2016}) and to the fact that only a small fraction of the thermal tails is detected in our data. 
In addition, different observing epochs imply different observing geometry, which can induce additional size and albedo uncertainty in the case of an elongated object. 
Taken together, these various sources of uncertainties can lead to a 50\% size uncertainty and a 100\% albedo uncertainty. 
This is reflected here in the case of NEOs (438429) 2006~WN1 and 2017~YE5, where the albedo values vary almost by a factor two between the two observing visits to these asteroids.


\begin{deluxetable}{llccc}[h!]
\tabletypesize{ \scriptsize}
\tablecaption{Thermal Parameters}
\tablehead{Number	&	Designation	&	$\eta$	&	${\rm p_v}$ &	Observing Date}
\startdata
52768	&	1998 OR2	&	1.232	&	0.128	&	28/02/2020	\\
90416	&	2003 YK118	&	1.611	&	0.017	&	18/02/2015	\\
163348	&	2002 NN4	&	0.931	&	0.150	&	17/06/2020	\\
275611	&	1999 XX262	&	0.928	&	0.014	&	06/02/2019	\\
354030	&	2001 RB18	&	1.039	&	0.019	&	26/09/2019	\\
410778	&	2009 FG19	&	1.390	&	0.008	&	26/09/2019	\\
438429	&	2006 WN1	&	1.314	&	0.013	&	20/07/2015	\\
438429	&	2006 WN1	&	1.142	&	0.025	&	03/09/2018	\\
452389	&	2002 NW16	&	1.461	&	0.017	&	15/08/2016	\\
471241	&	2011 BX18	&	1.308	&	0.031	&	15/08/2016	\\
	    &	2011 YS62	&	1.254	&	0.015	&	03/12/2015	\\
	    &	2014 UF206	&	1.059	&	0.055	&	18/01/2015	\\
	    &	2014 LW21	&	1.031	&	0.051	&	12/07/2020	\\
	    &	2015 SV2	&	1.079	&	0.035	&	12/12/2015	\\
	    &	2016 PR8	&	1.270	&	0.010	&	03/01/2017	\\
	    &	2016 XH	    &	0.992	&	0.038	&	24/12/2016	\\
	    &	2017 YE5	&	1.926	&	0.020	&	21/06/2018	\\
	    &	2017 YE5	&	1.661	&	0.012	&	23/06/2018	\\
	    &	2019 FU	    &	1.624	&	0.034	&	08/04/2019	\\
\enddata
\tablenotetext{}{$\eta$=beaming parameter, ${\rm p_v}$=visual (V-band) albedo.}
\label{tab:thermal}
\end{deluxetable}

\newpage

\section{Spectral comparison between NEOs and MBAs for all taxonomic classes}
\label{sec:app_D}

\begin{figure}[h!]
\centering
\includegraphics[angle=0, width=0.45\linewidth, trim=0cm 0cm 0cm 0cm, clip]{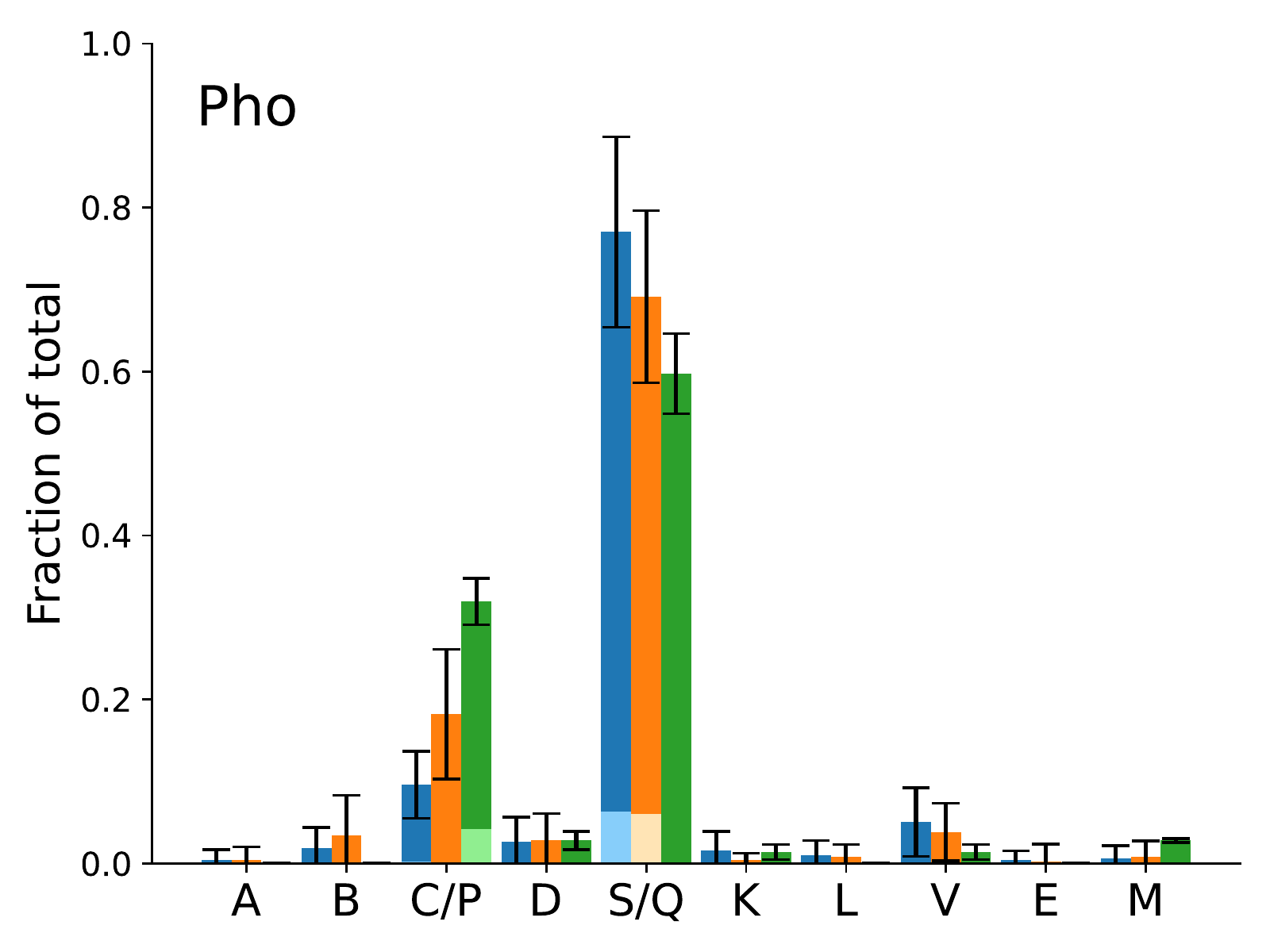}
\includegraphics[angle=0, width=0.45\linewidth, trim=0cm 0cm 0cm 0cm, clip]{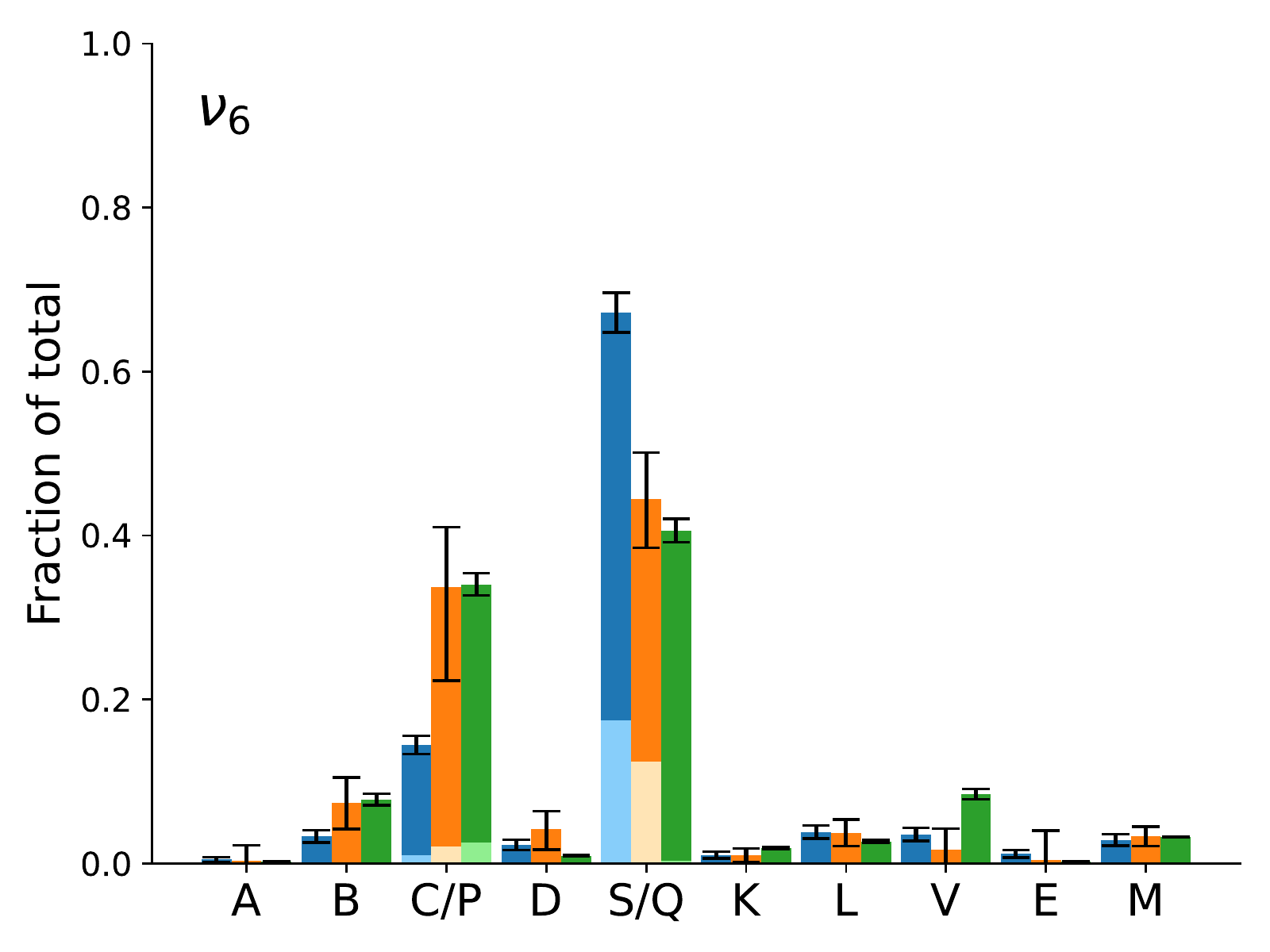}
\includegraphics[angle=0, width=0.45\linewidth, trim=0cm 0cm 0cm 0cm, clip]{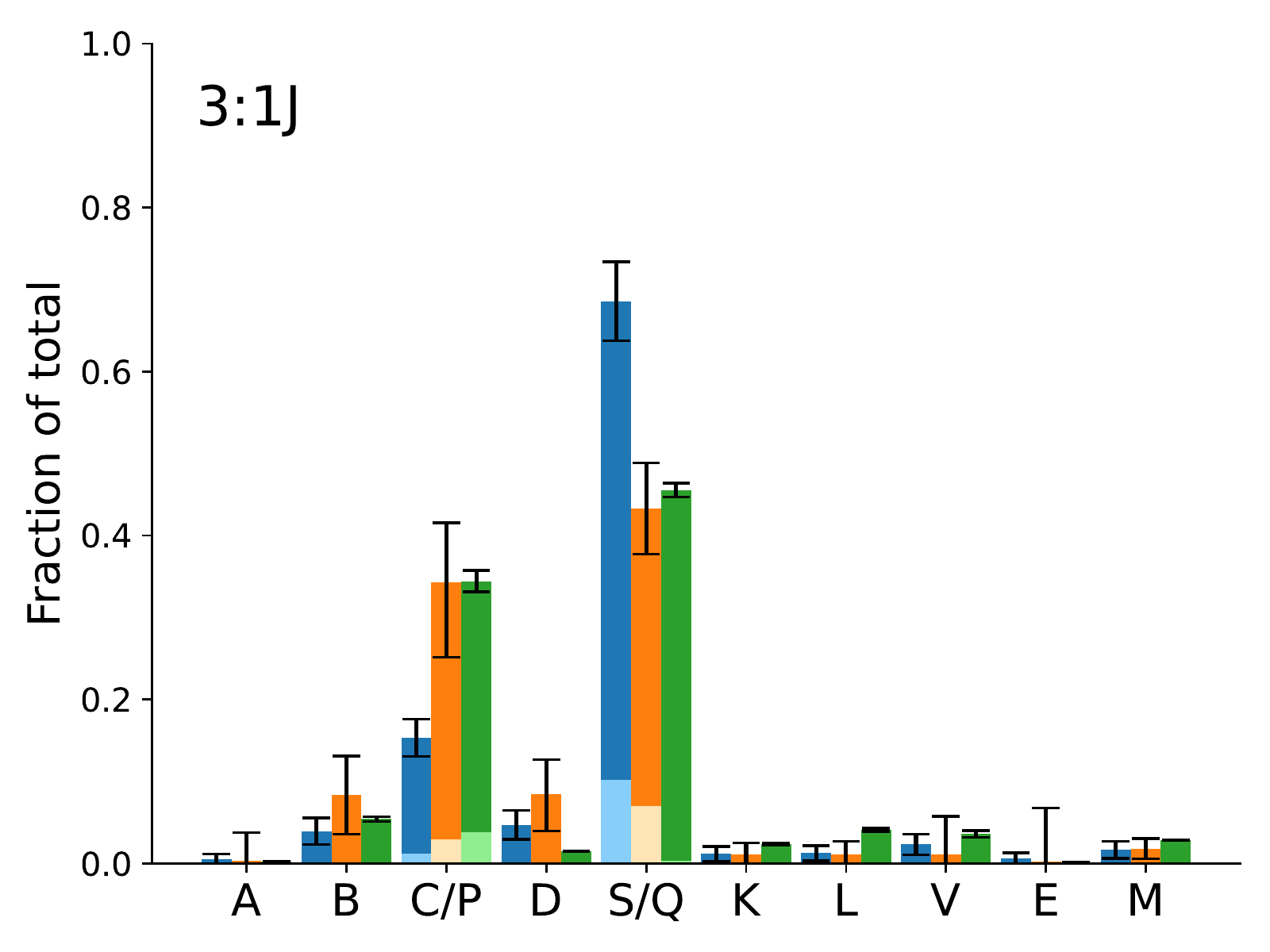}
\includegraphics[angle=0, width=0.45\linewidth, trim=0cm 0cm 0cm 0cm, clip]{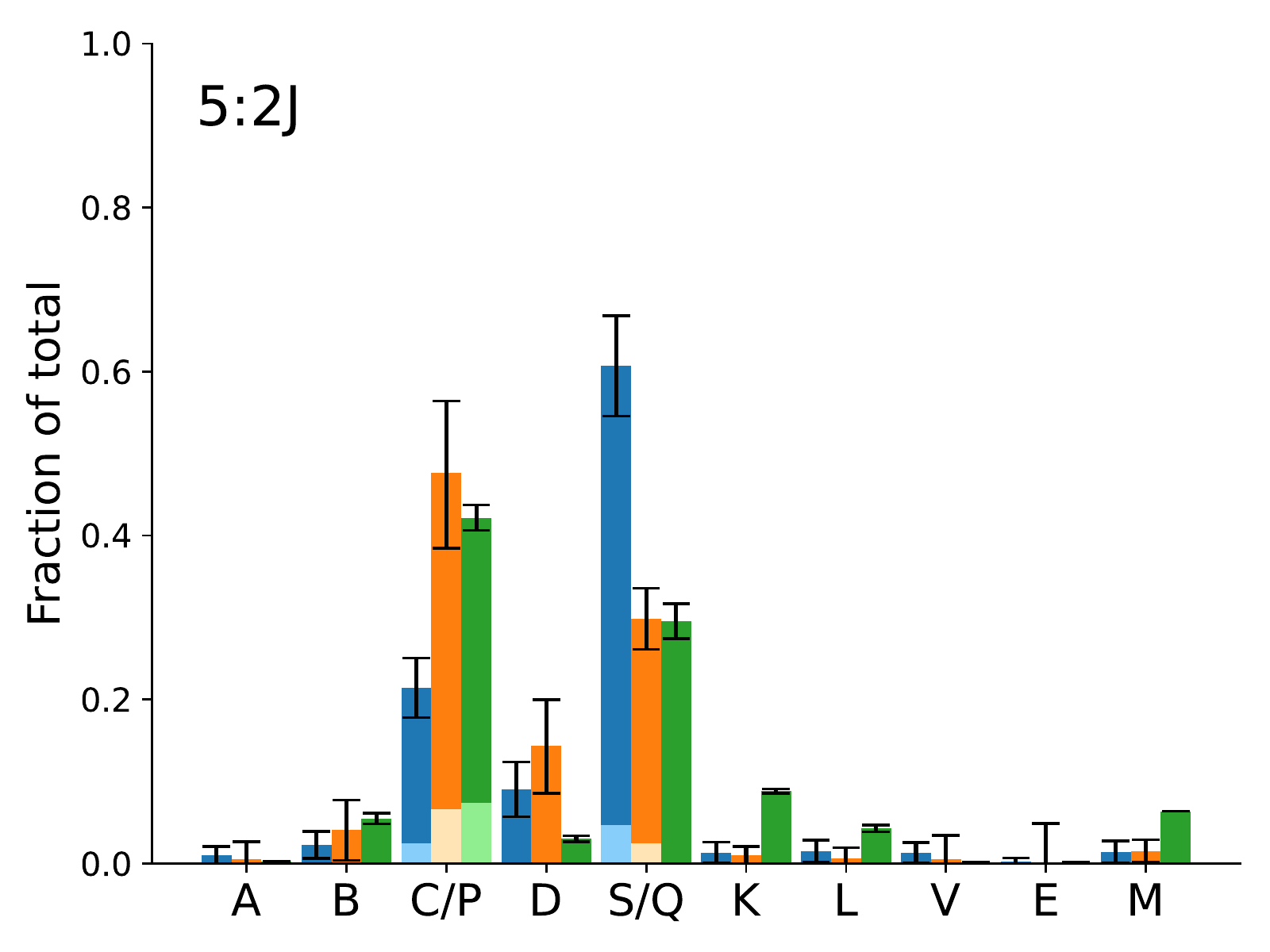}
\includegraphics[angle=0, width=0.45\linewidth, trim=0cm 0cm 0cm 0cm, clip]{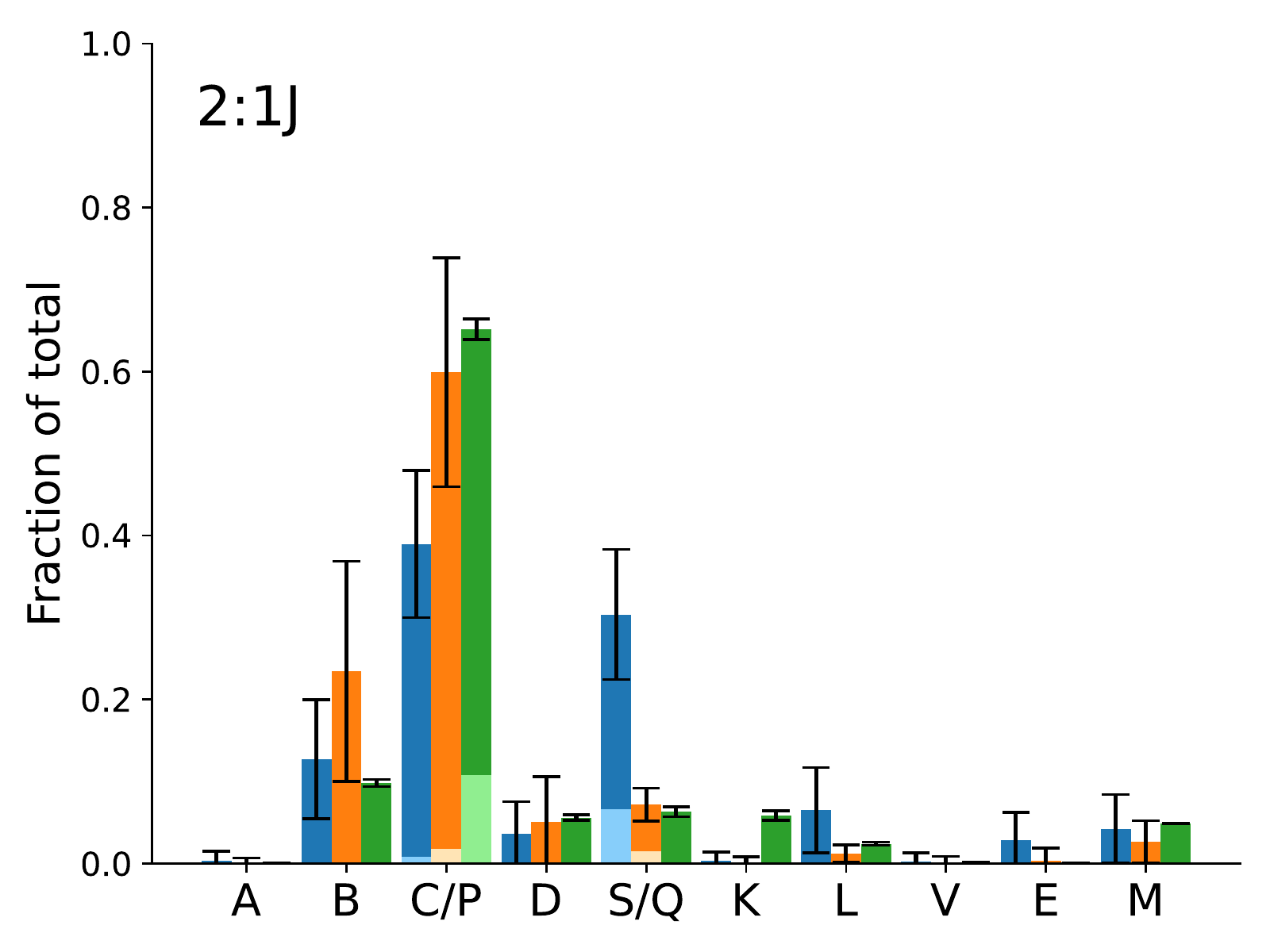}
\includegraphics[angle=0, width=0.45\linewidth, trim=0cm 0cm 0cm 0cm, clip]{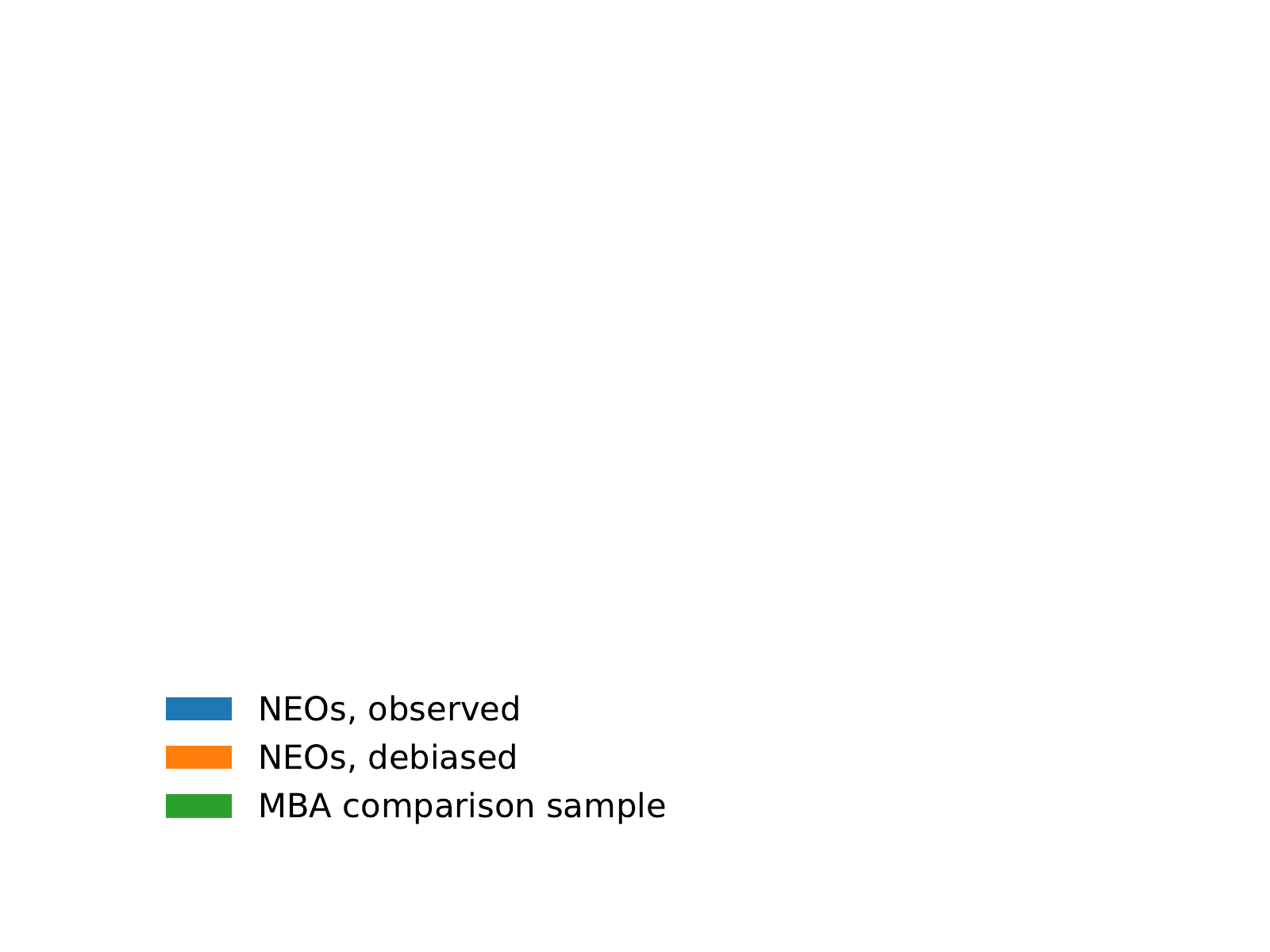}
 \caption{Same as Fig.\,\ref{fig:debAbundances_MBAalb}, with the inclusion of the minor (least-populated) taxonomic classes.}
\label{fig:debAbundances_MBAalb2}
\end{figure}

Comparison between the spectral class distributions of NEOs and MBAs for all taxonomic classes is presented in Figure \ref{fig:debAbundances_MBAalb2}.
\DIFaddend 
\DIFaddbegin 

\newpage

\section{Complete tabulation}
\label{sec:app_E}

A description of the complete set of observations, spectral analysis, NEO orbital parameters 
and escape region probabilities is provided in Table \ref{tab:suppmat}.
The complete dataset is available in a machine-readable format in the online Journal.

\begin{deluxetable*}{lll}[!hb]
\tablewidth{0pt}
\tablecolumns{3}
\tablecaption{Complete tabulation\label{tab:suppmat}}
\tablehead{%
\colhead{Label} & 
\colhead{Unit} & 
\colhead{Description} 
}
\startdata
File & --- & Spectrum filename \\ 
Number & --- & Near-Earth Object Number \\ 
Desig & --- & Provisional Designation \\ 
Name & --- & Name \\ 
a & au & Semi-major Axis \\ 
e & --- & Eccentricity \\ 
i & deg & Inclination \\ 
Hmag & mag & H Magnitude \\ 
TJup & --- & Jupiter Tisserand \\ 
Delta-v & km/s & Velocity Change Needed for Rendez-vous Mission \\ 
Hazard? & --- & Potentially Hazardous? \\ 
Taxon & --- & Taxonomy \\ 
Slope & um-1 & Spectral Slope \\ 
PC1' & --- & 1st Principal Component \\ 
PC2' & --- & 2nd Principal Component \\ 
PC3' & --- & 3rd Principal Component \\ 
PC4' & --- & 4th Principal Component \\ 
PC5' & --- & 5th Principal Component \\ 
ObsDate-S & --- & Start Date Time, ISO 8601 format \\ 
ObsDate-E & --- & End Date Time, ISO 8601 format \\ 
RAh & h & Hour of Right Ascension (J2000)  \\ 
RAm & min & Minute of Right Ascension (J2000)  \\ 
RAs & s & Second of Right Ascension (J2000)  \\ 
DE- & --- & Sign of the Declination (J2000) \\ 
DEd & deg & Degree of Declination (J2000)  \\ 
DEm & arcmin & Minute of Declination (J2000)  \\ 
DEs & arcsec & Second of Declination (J2000)  \\ 
AM-S & --- & Start Airmass \\ 
AM-E & --- & End Airmass \\ 
ParAng-S & deg & Start Parallactic Angle \\ 
ParAng-E & deg & End Parallactic Angle  \\ 
ExpTime & s & Exp. Time \\ 
Nimages & --- & Number of Images \\ 
TotalExp & s & Total Exp. Time \\ 
GeoDist & au & Geocentric Distance \\ 
SolDist & au & Solar Distance \\ 
PhaseAng & deg & Phase Angle \\ 
Vismag & mag & Vis. Magnitude \\ 
P[Hun] & --- & Hungaria Escape Route Probability \\ 
P[nu6] & --- & nu6 Escape Route Probability \\ 
P[Pho] & --- & Phocaea Escape Route Probability \\ 
P[3:1J] & --- & 3:1J Escape Route Probability \\ 
P[5:2J] & --- & 5:2J Escape Route Probability \\ 
P[2:1J] & --- & 2:1J Escape Route Probability \\ 
P[JFC] & --- & JFC Escape Route Probability \\ 
\enddata
\tablecomments{Principal components were calculated after removing the spectral slope from the asteroid spectrum.}
\tablecomments{The complete table is available in a machine-readable format in the online Journal.}
\end{deluxetable*}



\end{document}